\documentclass{article}
\usepackage[dvipdfmx]{graphicx}
\usepackage{amsmath}
\usepackage{amssymb}
\usepackage{authblk}
\usepackage{mathrsfs}
\usepackage{slashed}
\usepackage{color}
\usepackage{cite}
\usepackage{here}
\usepackage{subcaption}
\usepackage{ulem}

\makeatletter
 
 \@addtoreset{equation}{section}
 \makeatother

\newcommand{\bea}{\begin{equnarray}}
\newcommand{\eea}{\end{eqnarray}}
\newcommand{\beann}{\begin{equnarray*}}

\newcommand{\eeann}{\end{eqnarray*}}
\newcommand{\nn}{\nonumber}
\newcommand{\ba}{\begin{array}}
\newcommand{\ea}{\end{array}}

\newcommand{\bs}{\boldsymbol}

\newcommand{\bse}{{\boldsymbol{e}}}
\newcommand{\bsone}{\bs{1}}

\newcommand{\N}{\mathbb N}

\newcommand{\tC}{{\tilde{C}}}

\newcommand{\cB}{{\cal B}}

\newcommand\dashint{\mathchoice
  {\int\kern-10pt-}
  {\int\kern-8.5pt-}
  {\int\kern-6.1pt-}
  {\int\kern-4.58pt-}}

\DeclareMathOperator{\Tr}{Tr}

\title{Gross-Witten-Wadia Phase Transition in Induced QCD\\on the Graph} 
\author[1]{So Matsuura\thanks{s.matsu@keio.jp}}
\author[2]{Kazutoshi Ohta\thanks{kohta@law.meijigakuin.ac.jp}}
\affil[1]{\it Hiyoshi Departments of Physics,
and Research and Education Center for Natural Sciences,
Keio University, 
Yokohama, Kanagawa 223-8521, Japan}
\affil[2]{\it Institute of Physics, Meiji Gakuin University, Yokohama, Kanagawa 244-8539, Japan}

\date{}

\begin{document}
\maketitle

\vspace*{2cm}

\begin{center}
{\bf Abstract}
\end{center}

In this paper, we examine a modification of the Kazakov-Migdal (KM) model with gauge group $U(N_c)$, where the adjoint scalar fields in the conventional KM model  are replaced by $N_f$ fundamental scalar fields (FKM model). After tuning the coupling constants and eliminating the fundamental scalar fields, the partition function of this model is expressed as  an integral of a graph zeta function weighted by unitary matrices.
The FKM model on cycle graphs at large $N_c$ exhibits the Gross-Witten-Wadia (GWW) phase transition only when $N_f > N_c$. In the large $N_c$ limit, we evaluate the free energy of the model on a general graph in two distinct parameter regimes and demonstrate that the FKM model generally consists of multiple phases.
The effective action of the FKM model reduces to the standard Wilson action by taking an appropriate scaling limit when the graph consists of plaquettes (fundamental cycles) of the same size, as in the square lattice case. We show that, for the FKM model on such a graph, the third-order GWW phase transition universally occurs in this scaling limit.
\newpage

\section{Introduction}
\label{sec:Introduction}

Quantum Chromodynamics (QCD) is the quintessential theory that describes the dynamics of
gluons and quarks. 
In general, gauge invariant operators play a pivotal role in gauge theory,
but the Wilson loops are especially important among them. 
In fact, the Yang-Mills theory, which characterizes quantum gluodynamics (QGD), can be comprehended through the Wilson loops as fundamental degrees of freedom: The action of lattice QCD, a powerful formulation that presents QCD in a non-perturbative manner, is given by the sum of Wilson loops on the plaquettes \cite{PhysRevD.10.2445}. 
The Schwinger-Dyson equation that the Wilson loops obey forms a closed equation at large $N$ \cite{makeenko1979exact,wadia1981dyson}, 
which plays a major role in understanding the duality between large $N$ gauge theory and string theory.

The Kazakov-Migdal (KM) model is a gauge theory defined originally
on a $D$-dimensional square lattice as \cite{kazakov1993induced}
\begin{equation}
  S_{\rm KM}= N \sum_x \Tr\left\{m_0^2 \Phi^2(x) - \sum_{\mu=1}^D
  \Phi(x)U_\mu(x)\Phi(x+\mu) U_\mu^\dagger(x)
  \right\}\,,
  \label{eq:KM}
\end{equation}
where $U_\mu(x)$ is a unitary variable living on the link extending from the site (vertex) $x$ to the
{neighbor sites along the} 
direction of $\mu$ and
$\Phi(x)$ is a scalar field in the adjoint representation of the gauge group living on $x$.
The distinctive feature of the KM model is that the effective action resulting from the integration of the scalar fields is comprised of the sum of (the square of the absolute value of) 
possible Wilson loops on the lattice.
Therefore, this simple model was expected to induce QCD (QGD in precise) in arbitrary dimensions and was vigorously studied \cite{migdal19931, gocksch1992phase, khokhlachev1992problem, caselle1992exact, aoki1993spectrum, aoki1993study, makeenko1993exact}.
Despite early expectations of the KM model inducing QGD, however, 
it has become clear that the presence of an extra {\it local} $U(1)$ symmetry
\begin{equation}
U_\mu(x)\to e^{i\alpha_\mu(x)}U_\mu(x),
\label{eq:local U(1)}
\end{equation}
and the absence of the first-order phase transition preclude its induction of QGD \cite{kogan1992induced,kogan1993area,kogan1993continuum, migdal1993bose,cline1993induced,balakrishna1994difficulties}. 

Even though the objective of inducing QGD was not attained, the KM model, with its capability to be solved exactly at large $N$, remains a noteworthy subject of study. 
In \cite{boulatov1993infinite}, the correspondence between the large $N$ KM model and string theory was pointed out.
The work of \cite{caselle1993kazakov,caselle1994two} has demonstrated that the $D$-dimensional KM model serves as a description of the high-temperature limit of the $D+1$-dimensional Yang-Mills theory. Additional insights into the relationship between the KM model and the Penner model can be found in \cite{makeenko1993some,paniak1995kazakov}. 
More recently, in \cite{matsuura2022kazakov,matsuura2022graph}, the authors of this paper 
generalized the KM model to the one 
defined on an arbitrary graph and pointed out a relation to the graph zeta functions \cite{Ihara:original,MR607504,sunada1986functions,bartholdi2000counting,mizuno2003bartholdi,sato2006weighted,choe2007bartholdi}. 
Essentially, the graph zeta function is a function that enumerates cycles in a graph, 
and its compatibility with the Wilson loop counting provides a powerful tool for the analysis of KM models. 
As a result, in the large $N$ limit, the partition function of the KM model on arbitrary graphs can be directly evaluated exactly.

Despite the failure of the KM model to induce QGD being primarily attributed to its extra local symmetry, it remains a compelling endeavor to modify the model.
The first strategy attempted was to modify the theory so that the unwanted local symmetry is broken by a first-order phase transition and to expect QGD to be induced after the phase transition.
In \cite{khokhlachev1992adjoint}, 
the adjoint scalar of the KM model is replaced by adjoint fermions and it was suggested that the theory undergoes a first order phase transition. 
In \cite{ migdal1993mixed}, heavy fundamental fermions were added to the original KM model,
suggesting that there is a region that induces QGD. 
In \cite{ilchev1994mixed}, fundamental scalars were added to the original KM model, 
and numerical simulations showed that this model possesses a rich phase structure.

While these modifications are interesting in the sense that they leave open the possibility of inducing QGD, 
in this paper, 
we focus on the model proposed in \cite{Arefeva:1993ik} where QGD is realized {\it directly} by a simpler mechanism.
The model is obtained by replacing the adjoint scalar field in the original FKM model 
by complex scalar fields in the {\it fundamental} representation, 
which breaks the unwanted local symmetry explicitly.
In \cite{Arefeva:1993ik}, the author considered mainly on the case with the same number of the scalar fields with the rank of the gauge group. 
Therefore, 
the scalar fields are still $N\times N$ matrices,
but the action is modified to
\begin{equation}
  S' = N \sum_x \Tr\left\{m_0^2 \Phi^\dag(x)\Phi(x) - \sum_{\mu=1}^D
  \left[
  \Phi^\dag(x)U_\mu(x)\Phi(x+\mu)
  +\Phi^\dag(x+\mu)U^\dag_\mu(x)\Phi(x)
  \right]
  \right\}\,,
  \label{eq:FKM}
\end{equation}
which apparently does not have extra local $U(1)$ symmetry (\ref{eq:local U(1)}).
After integrating out the fields $\Phi$ and $\Phi^\dag$ by the Gaussian integral,
we obtain an effective action, 
\begin{equation}
  S'_{\rm eff} = N \sum_C
  \frac{1}{\ell(C) m_0^{2\ell(C)}}
  \Tr\left(
  U_C+U_C^\dag
  \right)\,,
  \label{eq:FKM induced}
\end{equation}
where $C$ and $\ell(C)$ represent all possible loops (cycles) on the $D$-dimensional lattice
and the ``naive'' length of $C$ which counts also the backtrackings,
and $U_C$ is the Wilson loop defined along $C$.
Since the possible loops $C$ on the lattice, of course,
contain the Wilson loop associated with each plaquette,
the effective action (\ref{eq:FKM induced}) contains the standard Wilson action
in the leading terms of the shortest loops. 
In order to make the model (\ref{eq:FKM}) equivalent to the Wilson action, a mechanism to suppress the longer loops is necessary. 
In \cite{Arefeva:1993ik}, it has been discussed that it is achieved by introducing the large number of the fundamental scalar fields 
with appropriately controlling the mass parameter. 
Since the model (\ref{eq:FKM}) has the same number of the fundamental scalar fields with the rank of the gauge group $N$, 
it has been expected to induce QGD by taking the limit of $N \to \infty$. 
The author then has mainly considered the model on a single plaquette in the limit of $N\to\infty$ to investigate a relation to the Eguchi-Kawai reduction.


However, there is a naive point in this scenario: 
The problem is that $\ell(C)$ does not represent the net length of the Wilson loop because the cycle $C$ contains backtrackings in general. 
Since there are infinitely many cycles with different $\ell(C)$ which reduce to the same Wilson loop, 
it is non-trivial if the long Wilson loops are really suppressed by simply increasing the number of the fundamental scalar fields. 
We have to estimate the coefficients of the Wilson loops accurately to confirm it. 
One of the purpose of this paper is to solve this problem by defining the model \eqref{eq:FKM} on an arbitrary graph and consider the number of the fundamental scalar fields $N_f$ as a parameter independent of the rank of the gauge group $N_c$. 
We call this model the FKM model in the following.
The FKM model does not have the extra $U(1)$ symmetry that the conventional KM model. 
A remarkable consequence of this generalization is that the effective action is expressed by an analytic function called the graph zeta function. 
In the previous papers of the present authors \cite{matsuura2022kazakov,matsuura2022graph}, the graph zeta function played a central role to count the Wilson loops of the original KM model, and we can use the same technology to analyze the FKM model.
In particular, it supplies a way to estimate the coefficients of the Wilson loops in the effective action. 
As a result, we can show that the effective action of the FKM model actually reduces to the Wilson action in an appropriate limit of the theory parameters.  
Furthermore, the FKM model in general parameters includes higher derivative terms arising from longer Wilson loops.
The use of graph zeta functions is a great advantage to be able to treat such general lattice gauge theories analytically%
\footnote{
We note that another possible model is obtained by replacing the fundamental scalars in the FKM model to fundamental fermions, which leads a sort of the Venetiano model \cite{veneziano1976some}. 
Of course, we cannot expect the same mechanism to realize QGD with the FKM model because the signature of the effective action becomes opposite to the Wilson action. 
However, it is still interesting approach to analyze general gauge theories since the introduction of the fundamental fermions is expected to produce a rich phase structure to the model in general \cite{tian1982phase}. 
We would like to thank the reviewer of PRD for pointing out this possibility. 
}.

Another important result of the present paper is
that the FKM model universally undergoes the Gross-Witten-Wadia (GWW) phase transition in large $N_c$ regardless of the details of the graph structure. 
The GWW phase transition was discovered as a large $N_c$ phase transition in the one-plaquette model obtained from two-dimensional lattice gauge theory \cite{Gross:1980he,Wadia:1980cp}, 
but is now recognized as a transition between disjoint/connected distribution of the eigenvalues of the Polyakov loops on the circumference, which is closely related to the confinement/deconfinement transition of QCD 
\cite{Hanada:2018zxn,Hanada:2019czd,hanada2020partial,Watanabe:2020ufk}
Interestingly, the original KM model does not enjoy the GWW phase transition. 
This is due to the inability of the attractive force between eigenvalues of the unitary matrix to surpass the repulsive force emanating from the measures, as the contribution of Wilson loops with finite area is suppressed in the limit of $N_c\to\infty$. 
The mechanism is the same in the FKM moddel: In order for the FKM model to undergo the GWW phase transition, the condition $N_f\gtrsim N_c$ must be satisfied 
for the attractive force from the action to exceed the repulsive force from the measure. 
This is a parameter region that had not been considered in the \cite{Arefeva:1993ik}.
Our analysis using the graph zeta function employed in \cite{matsuura2022kazakov,matsuura2022graph} proves useful even in this case, allowing us to rigorously evaluate the partition function of the FKM model on an arbitrary graph at large $N_c$. This analysis necessitates the utilization of different evaluation methods for $N_c\ll N_f$ and $q N_f\ll N_c$ with a coupling constant $q$, leading to dissimilar expressions for the free energy in each region. 
A close examination of this result provides evidence that the model possesses multiple phases in general. 
Considering that the FKM model includes the Wilson action as a limit, this suggests the usefulness of the FKM model (especially in the region $N_f>N_c$) as a model for studying QCD.

The organization of this paper is as follows:
In Sec.~2, we explain the terms of graph theory and define the FKM model on an arbitrary graph. 
We demonstrate that the partition function of the model can be expressed by the unitary matrix integrals of a matrix-weighted graph zeta function.
We also show that the effective action of the FKM model reduces to the Wilson action by taking an appropriate scaling limit.
In Sec.~3, we establish that the FKM model on cycle graphs experiences a third-order GWW phase transition when $N_f > N_c$.
In Sec.~4, we calculate the partition function of the FKM model on a general graph for large $N_c$ and discuss the phase structure of the FKM model on a general graph. 
We also show that the FKM model on the graph which consists of plaquettes of the same size universally undergoes the third-order GWW phase transition in an appropriate scaling limit.
Sec.~5 is dedicated to the conclusions and discussion.
In Appendix~\ref{app:zeta}, we summarize the definition and essential properties of the graph zeta functions to make the paper self-contained.

\section{Kazakov-Migdal model in the fundamental representation}
\label{sec:FKM}

\subsection{Notations in graph theory}

Let us consider a connected simple directed graph $G$. 
The set of vertices and edges are denoted as $V$ and $E$, respectively, with cardinalities $|V|=n_V$ and $|E|=n_E$. 
A directed edge is symbolized as a pair of vertices $e=\langle v,v'\rangle$,
where $v=s(e)$ and $v'=t(e)$ the ``source''  
and ``target'' of the edge arrow of $e$, respectively.
The inverse of an edge $e$ is defined as a reversed arrow edge, denoted by $e^{-1}=\langle v',v\rangle$.
The set of edges and their inverses are combined to form the set $E_D$, defined as $E_D = \{\bse_a|a=1,\cdots,2n_E\} \equiv \{e_1,\cdots,e_{n_E},e^{-1}_1,\cdots,e^{-1}_{n_E}\}$.

A path $P=(\bse_1,\cdots,\bse_k)$ $(\bse_a \in E_D)$ is a sequence of the edges that satisfies $t(\bse_a)=s(\bse_{a+1})$ $(a=1,\cdots,k-1)$,
where $k$ is called the length of the path $P$ which is expressed as $|P|$.
If two paths $P=(\bse_1,\cdots,\bse_k)$ and $P'=(\bse'_1,\cdots,\bse'_l)$ satisfy
$t(\bse_k)=s(\bse'_1)$, we can construct a new path of length $k+l$
by connecting as
$PP'\equiv(\bse_1\cdots,\bse_k,\bse'_1,\cdots,\bse'_l)$.
A backtracking in $P$ is two consecutive edges in $P$ such that $\bse_{a+1}^{-1}= \bse_a$.

When a path $C=(\bse_1,\cdots,\bse_k)$ satisfies $s(\bse_1)=t(\bse_k)$, $C$ is referred to as a cycle of length $k$, which is denoted as $|C|$.
A cycle $C$ is called primitive when $C$ does not satisfy $C \ne B^r$ for any cycle $B$ and $r\ge 2$.
A cycle $C=(\bse_1,\cdots,\bse_k)$ is called tailless when $\bse_k^{-1}\ne \bse_1$, which is equivalent to that $C^2$ has no backtracking.
Backtracking or tail is also called bump,
the number of bumps in a cycle $C$ is denoted by $b(C)$.

Two cycles $C=(\bse_1,\cdots,\bse_k)$ and $C'=(\bse'_1,\cdots,\bse'_k)$
are called equivalent when there exists an integer $m$ such that $\bse_a = \bse'_{a+m(\!\!\!\!\mod k)}$.
The equivalence class including cycle $C$ is denoted as $[C]$. A cycle $C$ is defined as reduced if it lacks both a backtracking and a tail. 
The set of representatives of the equivalence classes of all kinds of cycles containing bumps is denoted as $[{\mathcal P}]$, 
and we the set of representatives of reduced cycles is denoted as $[{\mathcal P}_R] \subset [{\mathcal P}]$.

Since a primitive reduced cycle $C$ has its inverse $C^{-1}$ also be a primitive reduced cycle of equal length, 
the set $[{\mathcal P}_R]$ can be partitioned into two disjoint unions;  
$[{\cal P}_R]=[\Pi_+] \sqcup [\Pi_-]$, 
where $[\Pi_-]$ consists of the inverses of elements in $[\Pi_+]$. These elements in $[\Pi_+]$ are referred to as (the representatives of) chiral primitive reduced cycles.

\subsection{Definition of the model}

We propose a model on a simple directed graph $G$, 
generalizing the model presented in \cite{Arefeva:1993ik}. 
Each edge $e\in E$ on the graph $G$ is assigned a unitary matrix $U_e \in U({N_c})$. 
$N_f$ complex scalar fields, denoted as $\Phi_{v I}=(\Phi_{v I})_i$ ($I=1,\cdots,N_f$, $i=1,\cdots,{N_c}$), 
reside on each vertex $v\in V$ and are in the fundamental representation of $U({N_c})$.
Note that these scalar fields can also be expressed as rectangular matrices of size ${N_c}\times N_f$, as $\Phi_v=(\Phi_v)_{iI}$.
Then we define a gauge theory on the graph $G$ by the action, 
\begin{align}
  S&= \sum_{v\in V} m_v^2 \Phi_{v}^{\dagger I} \Phi_{v I}
  -q\sum_{e\in E}\left( \Phi_{s(e)}^{\dagger\ \ I} U_e \Phi_{t(e) I} +  \Phi_{t(e)}^{\dagger\ \ I} U_e^\dagger \Phi_{s(e) I} \right) \,,
  \label{eq:action}
\end{align}
where $m_v^2$ is a mass parameter of the scalar fields $\Phi_v^I$ which is uniform for all $I\in\{1,\cdots,N_f\}$ 
and $q$ is a coupling constant.   
In contrast to the original model in \cite{Arefeva:1993ik}, where $m_v^2$ was fixed, we now adjust $m_v^2$ based on the relationship between the vertex $v$ and the surrounding edges, as 
\begin{equation}
  m_v^2 = 1-q^2(1-u)^2+q^2(1-u) \deg v\,,
  \label{eq:mass}
\end{equation}
where $\deg v$ is the number of edges whose source or target is the vertex $v$. 
This parametrization is necessary for the interpretation of the model's partition function as a graph zeta function in the following subsection, as established in \cite{matsuura2022kazakov,matsuura2022graph}.
While the model retains the $U(N_c)$ gauge symmetry, it lacks the extra $U(1)$ symmetry present in the original KM model \eqref{eq:KM}. 
Henceforth, we refer to this model as the FKM model. 

\subsection{Partition function in terms of the graph zeta function}

The integration over the scalar fields in the definition of the partition function of this model is a Gaussian integral; 
\begin{align}
  Z_G &= \int \prod_{v\in V}d\Phi_{v} d\Phi^{\dagger v} \prod_{e\in E}dU_e\,
  e^{-\beta S} \nn \\
  &= \prod_{I=1}^{N_f}\int \prod_{v\in V}d\Phi_{v I} d\Phi^{\dagger v I} \prod_{e\in E}dU_e
  \, e^{-\beta \Phi^{\dagger v I} \Delta_{v}^{\ v'} \Phi_{v' I}}\,, 
  \label{eq:Zmed}
\end{align}
where the action is given by \eqref{eq:action} 
and $\Delta$ is a square matrix of size ${N_c}n_V$ with the elements defined by
\begin{equation}
  \Delta_{v}^{\ v'} 
  \equiv \delta_{v}^{\ v'}\bsone_{N_c} - q (A_U)_{v}^{\ v'} + (1-u)q^2\left(
  \deg v  - (1-u)\right)\delta_{v}^{\ v'}\bsone_{N_c}\,, 
  \label{eq:matDelta}
\end{equation}
and the matrix $A_U$ is the matrix-weighted adjacency matrix whose elements are given by
\begin{equation}
  (A_U)_{v}^{\ v'} = \sum_{\bse\in E_D} U_{\bse}\,
   \delta_{\langle{v},{v'}\rangle,\bse} \,.
   \label{eq:AU}
\end{equation}
Although $\beta$ can be absorbed into the scalar fields by rescaling, it is deliberately retained in the expression.
The matrix $\Delta$ is the same matrix appearing in the vertex representation of the matrix-weighted Bartholdi zeta function\cite{matsuura2022kazakov,matsuura2022graph}, 
\begin{equation}
  \zeta_G(q,u;U) = \bigl(1-(1-u)^2q^2\bigr)^{-{N_c}(n_E-n_V)}
  \det\bigl(\bsone_{{N_c}n_V}-q A_U + (1-u)q^2(D-(1-u) \bsone_{N_c n_V})\bigr)^{-1}\,,
  \label{eq:vertex U-Bartholdi}
\end{equation}
where $D$ is a diagonal matrix whose elements are given by $D_{vi}^{\ v'j}=\deg v\, \delta_{v}^{\ v'}\delta_{i}^{\ j}$. 
Therefore, by integrating out the scalar fields, the partition function \eqref{eq:Zmed} can be written as
\begin{align}
  Z_G &= \left(\frac{2\pi}{\beta}\right)^{{N_f}{N_c}n_V}
  \left(1-(1-u)^2q^2\right)^{{N_f}{N_c}(n_E-n_V)}
  \int\prod_{e\in E}dU_e\,
  \zeta_G(q,u;U)^{N_f}\,.
  \label{eq:Zpre}
\end{align}
The matrix-weighted Bartholdi zeta function $\zeta_G(q,u;U)$ is originally defined as the Euler product over all possible Wilson loops along primitive cycles of the graph: 
\begin{align}
  \zeta_G(q,u;U) &\equiv \prod_{C\in [{\cal P}]} \det\bigl( \bsone_{{N_c}} - q^{|C|} u^{b(C)} U_C \bigr)^{-1}\,,
  \label{eq:U-Bartholdi}
\end{align}
where $U_C$ is the Wilson loop associated with the cycle $C=(\bse_{i_1}\cdots\bse_{i_{|C|}})$ and 
$U_C \equiv U_{\bse_{i_1}}\cdots U_{\bse_{i_{|C|}}}$.
If the cycle $C$ contains a bump $\bse \bse^{-1}$, the unitary matrices at the bump in $U_C$ are canceled as $U_{\bse}U_{\bse^{-1}}=U_{\bse}U_{\bse}^{-1}=\bsone_{N_c}$. 
Therefore, the Wilson loop on such a cycle is equivalent to the Wilson loop on the reduced cycle obtained by recursively collapsing all bumps in $C$, and the Euler product in \eqref{eq:U-Bartholdi} is rewritten as a product over the equivalence class of chiral primitive reduced cycles as (for derivation, see \cite{matsuura2022kazakov}) 
\begin{align}
  \zeta_G(q,u;U) &= {\cal V}_G(q,u)^{N_c} \prod_{C\in [\Pi_+]}
  \exp\left(
    \sum_{n=1}^\infty \frac{1}{n}{f_{C}(q,u)^n} 
    \left( 
       \Tr U_C^{n} + \Tr U_C^{\dagger\, n}  
    \right)
   \right)\,,
  \label{eq:path U-Bartholdi}
\end{align}
where ${\cal V}_G(q,u)$ and $f_C(q,u)$ are defined by
\begin{align}
{\cal V}_G(q,u) 
&\equiv \prod_{\tC\in [{\cal B}_0]} \frac{1}{1-q^{|\tC|}u^{b(\tC)}}, \quad
  f_{C}(q,u) \equiv \sum_{\tC\in [\cB(C)]} q^{|\tC|}u^{b(\tC)}\,,
  \label{eq:VG and FC}
\end{align}
respectively. 
$[{\cal B}_0]$ and $[{\cal B}(C)]$ are the sets of the representatives of primitive cycles that reduce to a point and a primitive reduced cycle $C$, respectively, by eliminating all bumps. 
Combining \eqref{eq:Zpre} and \eqref{eq:path U-Bartholdi}, 
we can write the partition function as
\begin{align}
  Z_G &= \left(\frac{2\pi}{\beta}\right)^{{N_f}{N_c}n_V}
  \left(1-(1-u)^2q^2\right)^{{N_f}{N_c}(n_E-n_V)}
  {\cal V}_G(q,u)^{N_fN_c} \nn \\
  &\qquad \times \int\prod_{e\in E}dU_e\,
  \prod_{C\in [\Pi_+]}
  \exp\left(
    N_f \sum_{n=1}^\infty \frac{1}{n}{f_{C}(q,u)^n} 
    \left( 
       \Tr U_C^{n} + \Tr U_C^{\dagger\, n}  
    \right)
   \right)\,.
  \label{eq:Z}
\end{align}
Therefore, the effective action of the FKM model obtained by integrating the scalar field is 
\begin{equation}
    S_{\rm eff}(U) = 
  -\sum_{C\in [\Pi_+]}
    N_f \sum_{n=1}^\infty \frac{1}{n}{f_{C}(q,u)^n} 
    \left( 
       \Tr U_C^{n} + \Tr U_C^{\dagger\, n}
    \right)\,.
    \label{eq:Seff}
\end{equation}

It would be worthwhile to mention the existence of a limit in which this effective action becomes the Wilson action. 
To this end, we recall that, for a primitive reduced cycle $C$, the cycle with the shortest length in $[B(C)]$ is $[C]$ because bumps generally increase the length of a cycle. 
Therefore, from the definition \eqref{eq:VG and FC}, the function $f_C(q,u)$ behaves as 
\begin{equation}
    f_C(q,u) \to q^{|C|}\,, 
\end{equation}
if $q$ is sufficiently small  regardless of the value of $u$. 
Here, let $l$ be the minimum length of the primitive reduced cycles of the graph.
Then, by setting, 
\begin{equation}
\gamma \equiv \frac{N_f}{N_c}\,, 
  \label{eq:def gamma}
\end{equation}
and taking the limit,
\begin{equation}
    q\to 0,\quad \gamma\to\infty,\quad \lambda\equiv \frac{1}{\gamma q^l}\text{ : fixed}\,, 
    \label{eq:limit}
\end{equation}
the effective action \eqref{eq:Seff} reduces to
\begin{equation}
    S_{\rm eff}(U) = 
    -\frac{N_c}{\lambda}  
  \sum_{C\in [\Pi_+^l]}
    \left( 
       \Tr U_C + \Tr U_C^\dagger 
    \right)\,,
    \label{eq:SWilson}
\end{equation}
where $[\Pi_+^l]$ is the set of representatives of chiral primary reduced cycles of the minimal length $l$. 
When the graph consists of the fundamental cycles (plaquettes) 
 of the same length (size) $l$ as the standard square lattice, 
this is nothing but the Wilson action
with the 't\,Hooft coupling $\lambda=N_c g^2$. 
This indicates that the FKM model reproduces the lattice gauge theory with the Wilson action at least classically%
\footnote{
Although this limit has already been mentioned in \cite{Arefeva:1993ik}, 
the capacity for precise evaluation in the limit has become attainable due to the expression of the partition function in terms of the graph zeta function.
}.

For the latter purpose, we note that the matrix-weighted Bartholdi zeta function \eqref{eq:U-Bartholdi} can also be expressed as
\begin{align}
\zeta_G(q,u;U)
&= \det\left( \bsone_{2N_c n_E} - q(W_U + u J_U)\right)^{-1}\,,
\label{eq:edge U-Bartholdi}
\end{align}
where the matrices $W_U$ and $J_U$ are referred to as the edge adjacency matrix and bump matrix, respectively, and are defined as
\begin{align}
  (W_U)_{\bse\bse'} = \begin{cases}
  U_{\bse} & {\rm if}\ t(\bse) = s(\bse')\ {\rm and}\ \bse'^{-1}\ne \bse \\
    0 & {\rm others}
  \end{cases}\,,\quad
  (J_U)_{\bse\bse'} = \begin{cases}
    U_{\bse} & {\rm if}\ \bse'^{-1}= \bse \\
    0 & {\rm others}
  \end{cases}\,,
  \label{eq:WJu}
\end{align}
respectively \cite{matsuura2022graph}. 

Setting $N_c=1$ and $U_e=1$ in the expressions \eqref{eq:vertex U-Bartholdi}, \eqref{eq:U-Bartholdi} and \eqref{eq:edge U-Bartholdi} results in the standard Bartholdi zeta function $\zeta_G(q,u)$ \cite{bartholdi2000counting}, 
\begin{align}
    \zeta_G(q,u) 
  &= \prod_{C\in[{\cal P}]} (1-q^{|C|}u^{b(C)})^{-1} \nn \\
  &= \bigl(1-(1-u)^2q^2\bigr)^{-(n_E-n_V)}
  \det\bigl(\bsone_{n_V}-q A + (1-u)q^2(D-(1-u) \bsone_{n_V})\bigr)^{-1} \nn \\
  &= \det\bigl( \bsone_{2n_E} - q(W + u J)\bigr)^{-1}\,,
  \label{eq:Bartholdi zeta}
\end{align}
where the matrices $A$, $W$ and $J$ are obtained by setting $N_c=1$ and $U_e=1$ for $A_U$, $W_U$ and $J_U$ in \eqref{eq:AU} and \eqref{eq:WJu}, respectively. 
Furthermore, taking $u=0$ yields the Ihara zeta function \cite{Ihara:original,MR607504,sunada1986functions}, 
\begin{align}
    \zeta_G(q) 
  &= \prod_{C\in[{\cal P}_R]} (1-q^{|C|})^{-1} \nn \\
  &= \bigl(1-q^2\bigr)^{-(n_E-n_V)}
  \det\bigl(\bsone_{n_V}-q A + q^2(D-\bsone_{n_V})\bigr)^{-1} \nn \\
    &= \det\bigl( \bsone_{2n_E} - qW\bigr)^{-1}\,. 
  \label{eq:Ihara zeta}
\end{align}
These expressions will be also utilized later on.

\section{GWW phase transition in the FKM model on cycle graphs}
\label{sec:GWW}

\begin{figure}[H]
  \begin{center}
  \includegraphics[scale=0.5]{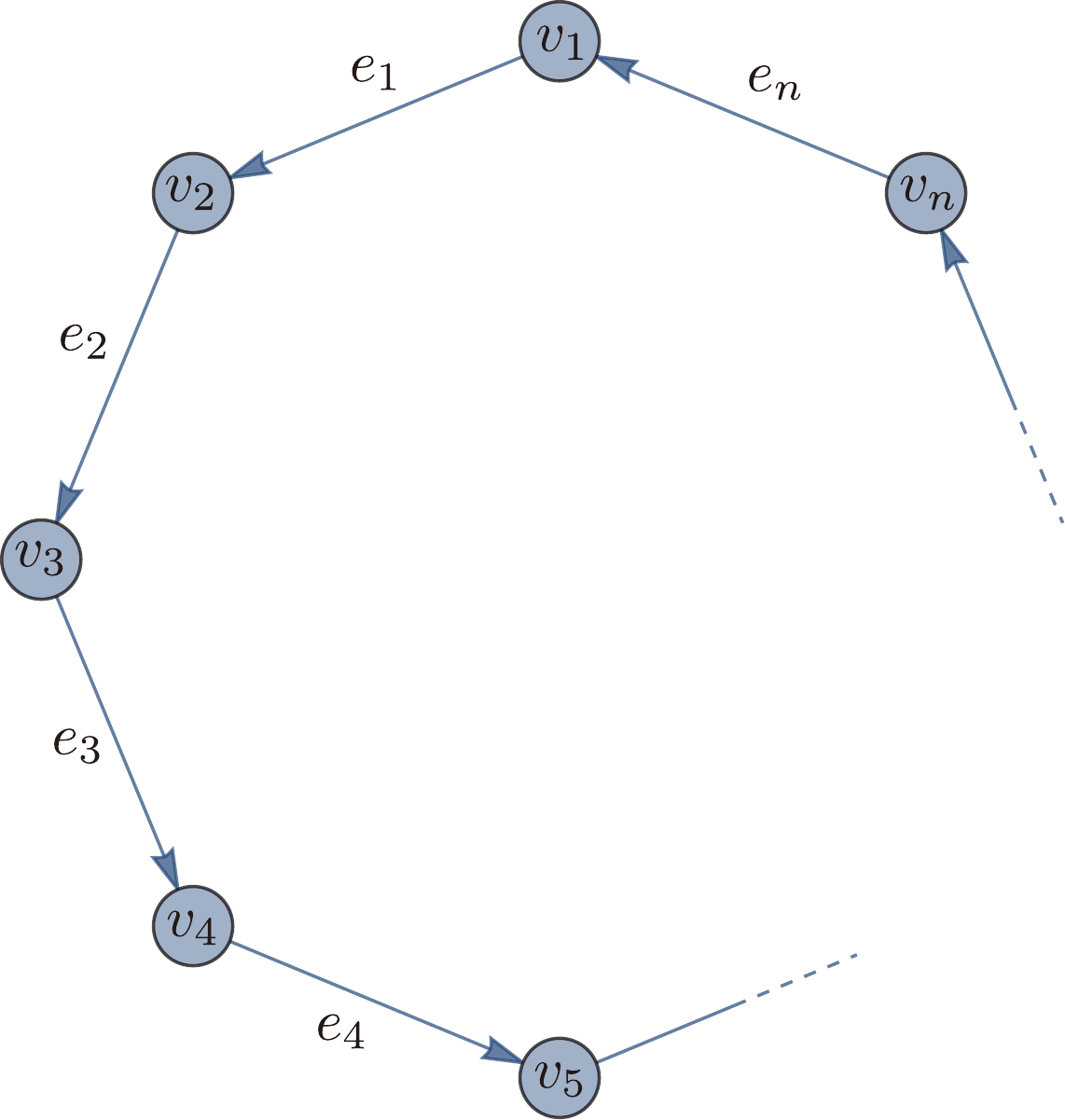}
  \end{center}
  \caption{A cycle graph $C_n$ with $n$ vertices and edges.}
  \label{fig:Cn}
\end{figure}
Let us examine the FKM model on a cycle graph $G=C_n$, which has $n_V=n_E=n$ and $\deg v=2$ for all vertices as depicted in Fig.~\ref{fig:Cn}. The cycle graph has a single chiral primitive reduced cycle $C=(e_1\cdots e_n)$, 
so we can simplify the notation for $U_C$ to $U$. In this instance, we can calculate ${\cal V}_G(q,u)$ and $f_C(q,u)$ explicitly as \cite{matsuura2022graph}
\begin{align}
  {\cal V}_G(q,u) = \xi_+(q,u)^{-n}, \quad
  f_C(q,u) = q^n \xi_+(q,u)^{-n}\,,
\end{align}
where 
\begin{equation}
  \xi_+(q,u) \equiv \frac{1}{2}\Bigl( 1+(1-u^2)q^2 + \sqrt{1-2(1+u^2)q^2 + (1-u^2)^2 q^4} \Bigr)\,.
\end{equation}
We assume $0<q<1$ and $0\le u<1$, 
in which region the Bartholdi zeta function on the cycle graph is well-defined. 
We define
\begin{equation}
\alpha \equiv f_C(q,u) = q^n\xi_+(q,u)^{-n}\,,
\end{equation}
which 
runs in the range $0<\alpha \le 1$. 
By fixing the gauge as $U_2=U_3=\cdots U_n=\bsone_{N_c}$, the partition function of the FKM model can be expressed as 
\begin{align}
  Z_{C_n} &=  \left(\frac{2\pi}{\beta q^n}\right)^{{N_f}{N_c}n_V}
  \alpha^{N_cN_f} \int dU e^{N_f\sum_{m=1}^\infty \frac{\alpha^m}{m} \left(\Tr U^m + \Tr U^{-m}\right)} \nn \\
  &= {\cal N} \int_{-\pi}^{\pi} \prod_{i=1}^{N_c} d\theta_i\,
  e^{\sum_{j\ne k}\log\left|\sin \frac{\theta_j-\theta_k}{2}\right|
   -N_f\sum_{i} \log\left(1-2\alpha \cos\theta_i + \alpha^2 \right)
  }\,,
  \label{eq:ZCn}
\end{align}
where ${\cal N}$ is an irrelevant normalization constant.

Let us next evaluate the free energy of this system 
in the limit of $N_c\to\infty$ and $N_f\to\infty$ with fixing the ratio 
$\gamma$ as in \eqref{eq:def gamma}, 
by following the strategy in \cite{Gross:1980he,Wadia:1980cp}.
In this limit, the eigenvalues $\theta_i$ can be treated as a continuous variable $\theta(x)$ $(x\in[0,1])$, and the free energy can be expressed as
\begin{align}
  F_{C_n} &\equiv -\lim_{N_c\to\infty} \frac{1}{N_c^2}\log Z_{C_n} \nn \\
  &= -{\dashint}_{\!\! 0}^{\,1}dxdy \log\left|\sin \frac{\theta(x)-\theta(y)}{2}\right|
  + \gamma\int_0^1 dx \, \log(1-2\alpha\cos\theta(x)+\alpha^2)\,,
  \label{eq:E0}
\end{align}
up to an inessential constant, where the symbol $\dashint$ denotes the principal value
integral. 

\begin{figure}[H]
\begin{center}
\subcaptionbox{$\alpha=0.01$}[.45\textwidth]{
\includegraphics[scale=0.45]{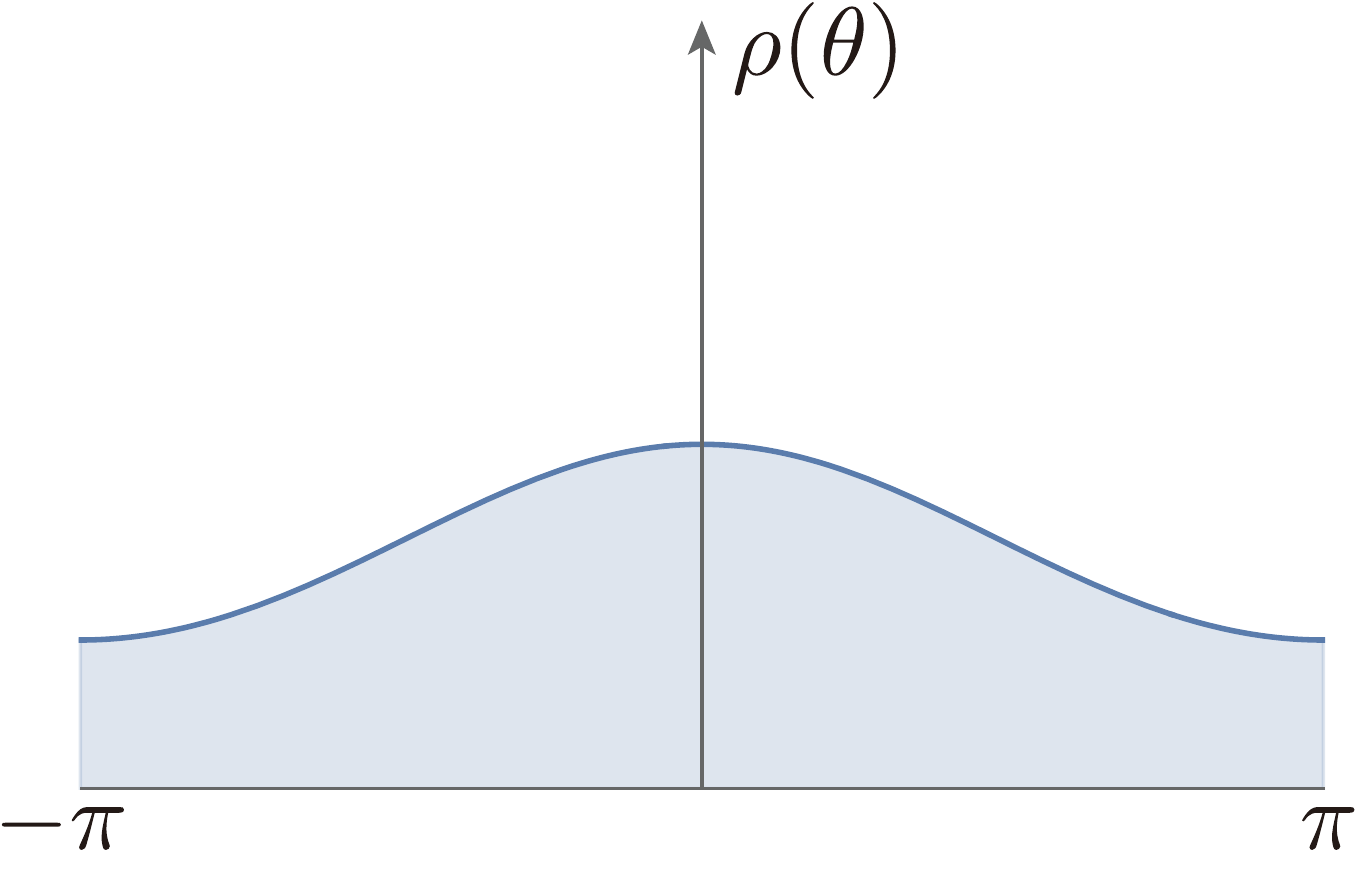}
}
\hspace*{0.5cm}
\subcaptionbox{$\alpha=0.08$}[.45\textwidth]{
\includegraphics[scale=0.45]{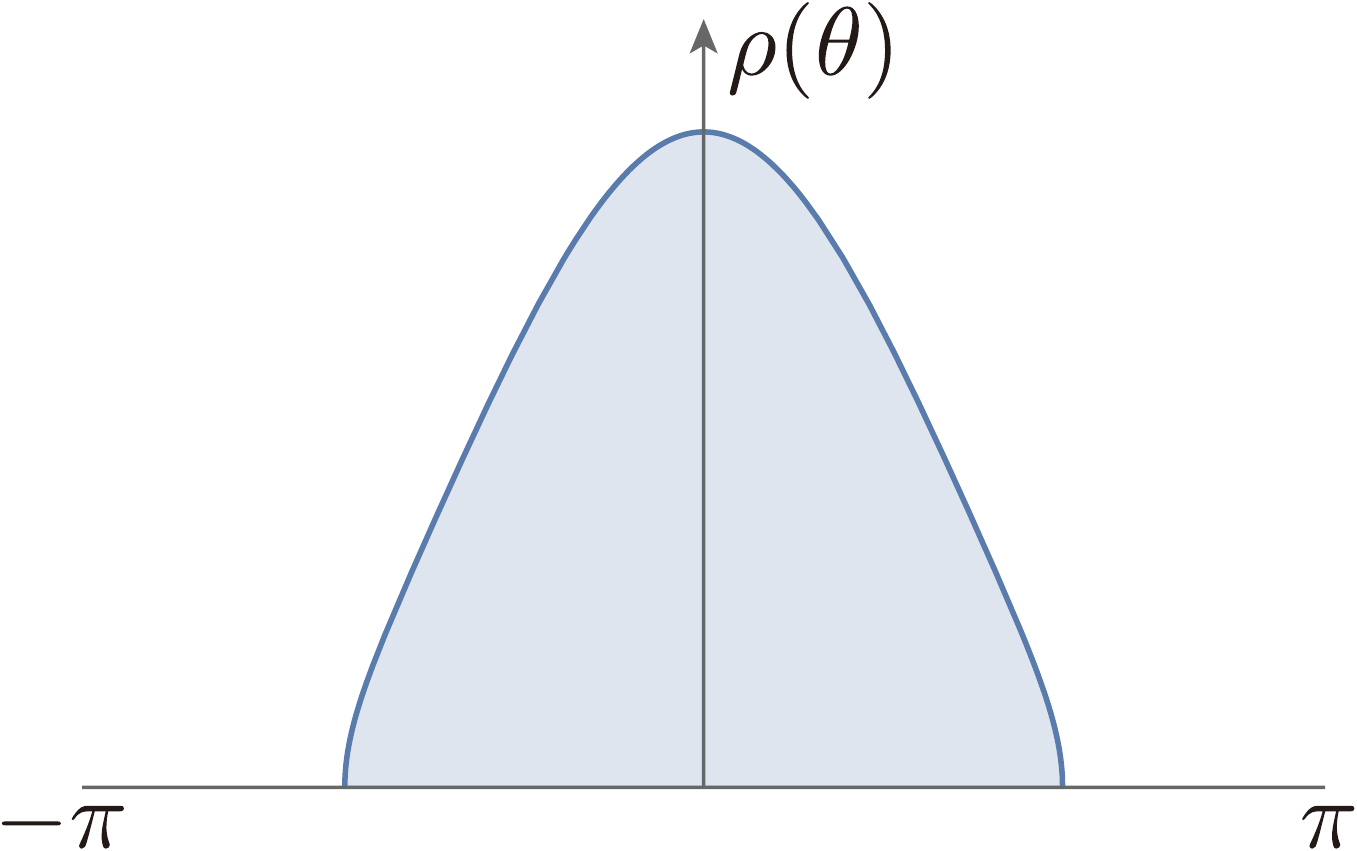}
}
\end{center}
\caption{
This show that an example of the eigenvalue distributions in the different phases.
We set $\gamma=10$, then the critical point is $\alpha^*=\frac{1}{19}=0.0526316\cdots$. 
(a)
For $\alpha<\alpha^*$, the eigenvalue distribution is attached at $\theta=\pm\pi$
with the periodicity. 
(b)
For $\alpha>\alpha^*$, the eigenvalue distribution
behaves in the same manner as Wigner's semi-circle.}
\label{Distribution of Eigenvalue}
\end{figure}

Since the saddle point approximation becomes exact in large $N_c$, the free energy is determined by solving the saddle point equation, 
\begin{equation}
  \dashint dy \, \cot\left(\frac{\theta(x)-\theta(y)}{2} \right)
  = {\dashint}_{\!\!\!-\theta_0}^{\,\theta_0} d\theta'\, \rho(\theta') \cot\left(\frac{\theta-\theta'}{2}\right)
  = \gamma \frac{2\alpha\sin\theta}{1-2\alpha\cos\theta+\alpha^2}\,,
  \label{eq:s.p.e}
\end{equation}
where $\rho(\theta)$ is the density of the eigenvalues, 
\begin{equation}
  \rho(\theta)\equiv \frac{1}{N_c}\sum_{i=1}^{N_c}\delta(\theta-\theta_i)\,,
\end{equation}
and $\pm\theta_0$ ($0<\theta_0\le\pi$) are the boundaries of the support of the density function, namely, we have assumed that $\rho(\theta)\ge 0$ in the region $-\theta_0\le \theta \le \theta_0$ and $\rho(\theta)=0$ outside.
If $\theta_0=\pi$, the eigenvalues are dispersed around the circle, whereas if $\theta_0<\pi$, the eigenvalues are distributed over only a portion of the circle around $\theta=0$
(Fig.~\ref{Distribution of Eigenvalue}). 
Under either boundary condition, the equation is easily solved \cite{jurkiewicz1983phase} and the result is 
\begin{equation}
  \rho(\theta) = \begin{cases}\displaystyle
    \frac{1}{2\pi}\left(1+2\gamma\frac{\alpha\cos\theta-\alpha^2}{1-2\alpha\cos\theta+\alpha^2}\right)\,, & \quad (\theta_0=\pi) \vspace{4mm}\\
    \displaystyle
    \frac{2(\gamma-1)\alpha}{\pi}\frac{\cos\frac{\theta}{2}}{1-2\alpha\cos\theta+\alpha^2}\sqrt{\sin^2\frac{\theta_0}{2}-\sin^2\frac{\theta}{2}}\,, & \quad (\theta_0<\pi)
  \end{cases}
  \label{eq:sln}
\end{equation}
where 
\begin{equation}
\sin^2\frac{\theta_0}{2} = \frac{(1-\alpha)^2}{4\alpha}\frac{2\gamma-1}{(\gamma-1)^2}\,.
\end{equation}
The phase transition between these two phases is known as the GWW phase transition \cite{Gross:1980he,Wadia:1980cp}. 

We should here note that the occurrence of the GWW phase transition in this model
depends on the value of $\gamma$: 
If $\gamma$ is too small, 
 the attractive force between the eigenvalues, induced by the potential (the second term of \eqref{eq:E0}), will be unable to adequately counteract the repulsive force caused by the Jacobian term (the first term), resulting in a circumferential distribution of eigenvalues, regardless of the value of $\alpha$.
In fact, in the phase of $\theta_0=\pi$, the condition $\rho(\pm\pi)\ge 0$ suggests that $0<\alpha\le\frac{1}{2\gamma-1}$ for $\gamma\ge\frac{1}{2}$ and $\alpha>\frac{1}{2\gamma-1}$ for $\gamma<\frac{1}{2}$. 
Recalling the range of $\alpha$, $0<\alpha\le 1$, this means that $\rho(\pm\pi)=0$ can be realized at 
\begin{align}
\alpha=\alpha^*\equiv \frac{1}{2\gamma-1}\,,
\end{align}
for $\gamma\ge1$, 
while $\rho(\pm\pi)$ is always positive for $\gamma<1$. 
In particular, for $\gamma=1$, the value of the $\alpha^*$ is at the boundary of the range of $\alpha$. 
Therefore, the GWW phase transition occurs only when $\gamma > 1$, that is, $N_f > N_c$.

Let us evaluate the free energy in both phases and determine the degree of the phase transition.
This is achieved by substituting \eqref{eq:sln} into \eqref{eq:E0}. 
After straightforward computations, we obtain
\begin{equation}
  \begin{split}
    F_{C_n}=\begin{cases}
      F_{C_n}^-\equiv \gamma^2\log\left(1-\alpha^2\right) &\quad (0<\alpha\le\alpha^*) \\
      F_{C_n}^+\equiv (2\gamma-1)\log(1-\alpha)+\frac{1}{2}\log\alpha + f(\gamma) &\quad (\alpha^*<\alpha \le 1)
    \end{cases}
  \end{split}
  \label{eq:FCn}
\end{equation}
where $f(\gamma)$ is a function of $\gamma$ which is determined by the requirement that both expressions have the same value at $\alpha=\alpha^*$.
Since the first and the second derivative by $\alpha$ of $F_{C_n}^\pm$ at $\alpha=\alpha^*$ are equal;
\begin{equation*}
\frac{dF_{C_n}^-}{d\alpha}\bigr|_{\alpha=\alpha^*}= \frac{dF_{C_n}^+}{d\alpha}_{\alpha=\alpha^*} = \frac{-2\gamma^2+\gamma}{2(\gamma-1)}\, ,
\end{equation*}
and
\begin{equation*}
\frac{d^2F_{C_n}^-}{d\alpha^2}\bigr|_{\alpha=\alpha^*}= \frac{d^2F_{C_n}^+}{d\alpha^2}\bigr|_{\alpha=\alpha^*} = \frac{(-2\gamma^2+2\gamma-1)(2\gamma-1)^2}{4(\gamma-1)^2}
\, ,
\end{equation*}
while the third derivative jumps;
\begin{equation*}
\frac{d^3F_{C_n}^-}{d\alpha^3}\bigr|_{\alpha=\alpha^*}\ne \frac{d^3F_{C_n}^+}{d\alpha^3}\bigr|_{\alpha=\alpha^*}\, .
\end{equation*}
This phase transition is third-order, which is the same as the original GWW phase transition.

It is instructive to observe that, upon taking the limit \eqref{eq:limit} with $q^n=\alpha$, 
the density \eqref{eq:sln} and the free energy \eqref{eq:FCn} converge to those of the original 
GWW model as 
\begin{equation}
  \rho(\theta) = \begin{cases}\displaystyle
    \frac{1}{2\pi}\left(1+\frac{2}{\lambda}\cos\theta\right)\,, & \quad (\theta_0=\pi) \vspace{4mm}\\
    \displaystyle
    \frac{2}{\pi\lambda}\cos\frac{\theta}{2}\sqrt{\frac{\lambda}{2}-\sin^2\frac{\theta}{2}}\,, & \quad (\theta_0<\pi)
  \end{cases}
\end{equation}
and
\begin{equation}
  \begin{split}
    F_{C_n}=\begin{cases}
      F_{C_n}^-\equiv -\frac{1}{\lambda^2} &\quad (0<\alpha\le\alpha^*) \\
      F_{C_n}^+\equiv -\frac{2}{\lambda}-\frac{1}{2}\log\lambda + {\rm const.} &\quad (\alpha^*<\alpha<1)
    \end{cases}\,,
  \end{split}
  \label{eq:FCn-limit}
\end{equation}
respectively. 
This is consistent with the result that the effective action of the FKM model becomes that of the 
GWW model (represented by the Wilson action in a single plaquette) in the limit \eqref{eq:limit}. 

\section{Analysis at large ${N_c}$ for general graphs}
\label{sec:general graph}

In this section, we evaluate the partition function of the FKM model on a general graph $G$ at large $N_c$.

\subsection{Evaluation by a large $N$ decomposition of Wilson loops}
\label{sec:decomp}


Let us first rewrite the partition function \eqref{eq:Z} by using the identity, 
\begin{equation}
  \exp\left({\sum_{m=1}^\infty \frac{a_m}{m}x^m}\right) = \sum_{n=0}^\infty x^n \sum_{\lambda\vdash n} \frac{a_\lambda}{z_\lambda}\,,
\end{equation}
where $\lambda$ is a partition of $n$
expressed as 
$\lambda=(l_1^{m_1}\cdots l_k^{m_k})$ $(l_1>\cdots > l_k \ge 0)$ with $\sum_{i=1}^k m_il_i=n$, 
$a_\lambda\equiv \prod_{i=1}^k a_{l_i}^{m_i}$, 
and $z_\lambda\equiv \prod_{i=1}^k m_i!\, l_i^{m_i}$. 
Applying this identity to the expression in \eqref{eq:path U-Bartholdi}, we obtain 
\begin{equation}
  \exp\left(
  {N_f\sum_{m=1}^\infty \frac{1}{m}f_C(q,u)^m \Tr U_C^m} 
  \right)
  = \sum_{n_C=0}^\infty f_C(q,u)^{n_C} \sum_{\lambda_C\vdash n_C} \frac{N_f^{m_{\lambda_C}}}{z_{\lambda_C}} \Upsilon_{\lambda_C}(U_C)\,, 
\end{equation}
where we have defined $m_\lambda\equiv \sum_{i=1}^k m_i$ and 
\begin{equation}
  \Upsilon_{\lambda}(U)  \equiv \prod_{i=1}^k \Tr\left(U^{l_i}\right)^{m_i}\,.
\end{equation}
Therefore, the unitary matrix integral in the partition function \eqref{eq:Z} can be rewritten as 
\begin{align}
  \int\prod_{e\in E}&dU_e\,
  \prod_{C\in [\Pi_+]}
  \exp\left(
    N_f \sum_{n=1}^\infty \frac{1}{n}{f_{C}(q,u)^n} 
    \left( 
       \Tr U_C^{n} + \Tr  U_C^{\dagger n}  
    \right)
   \right) \nn \\
   &= 
  \int\prod_{e\in E}dU_e\, 
  \prod_{C\in [\Pi_+]} 
   \left(
   \sum_{n_C,m_C=0}^\infty \sum_{\lambda_C\vdash n_C}\sum_{\mu_C\vdash m_C}
   f_C(q,u)^{n_C+m_C} \frac{N_f^{m_{\lambda_C}+m_{\mu_C}}}{z_{\lambda_C}z_{\mu_C}}
  \Upsilon_{\lambda_C}(U_C) \Upsilon_{\mu_C}(U_C^\dagger)
   \right) \,. \nn \\
  \label{eq:tmp}
\end{align}

Although it is hard to carry out this unitary matrix integral in finite $N_c$, 
it is decomposed in large $N_c$ as
\cite{Kazakov:1983fn, KOSTOV1984191, OBRIEN1985621} (see also \cite{matsuura2022graph})
\begin{equation}
  \int\prod_{e\in E}dU_e\, 
  \left(
  \prod_{C\in [\Pi_+]} \Upsilon_{\lambda_C}(U_C) \Upsilon_{\mu_C}(U_C^\dagger)
  \right)
  = 
  \prod_{C\in [\Pi_+]}\left(
  \int\prod_{e\in E}dU_e\, 
  \Upsilon_{\lambda_C}(U_C) \Upsilon_{\mu_C}(U_C^\dagger)
\right) +{\cal O}(1/N_c)\,,
\end{equation}
and the integral can be evaluated as 
\cite{matsuura2022kazakov} 
\begin{equation}
  \int\prod_{e\in E}dU_e\, 
  \Upsilon_{\lambda_C}(U_C) \Upsilon_{\mu_C}(U_C^\dagger) 
  = \delta_{m_C,n_C}\delta_{\lambda_C,\mu_C} z_{\lambda_C}+{\cal O}(1/N_c)\,,
\end{equation}
in large $N_c$.
Combining them, the integration \eqref{eq:tmp} can be rewritten as 
\begin{align}
  \int\prod_{e\in E}dU_e\,
  &\prod_{C\in [\Pi_+]}
  \exp\left(
    \sum_{n=1}^\infty \frac{1}{n}{f_{C}(q,u)^n} 
    \left( 
       \Tr U_C^{n} + \Tr  U_C^{\dagger n} 
    \right)
   \right) \nn \\
   &=  
   \prod_{C\in [\Pi_+]} \left(
   \sum_{n_C=0}^\infty f_C(q,u)^{2n_C} \sum_{\lambda_C\vdash n_C} \frac{N_f^{2m_{\lambda_C}}}{z_{\lambda_C}} 
   \right)
   (1+{\cal O}(1/N_c)) \nn \\
   &= 
   \left(
   \prod_{C\in [\Pi_+]} 
   \frac{1}{(1-f_C(q,u)^2)^{N_f^2}}
   \right)
   (1+{\cal O}(1/N_c))\,.
   \label{eq:tmp2}
\end{align}
In deriving the third line from the second line, we have used the identity, 
\begin{equation}
  \frac{1}{(1-t)^M} = \sum_{n=0}^\infty t^n \sum_{\lambda\vdash n}\frac{M^{m_\lambda}}{z_\lambda}\,,
  \label{eq:idpower}
\end{equation}
for $M\in\N$. 
This is derived from the generating function of the power sum symmetric polynomials,  
\begin{equation}
  \prod_{i} \frac{1}{1-tx_i} = \sum_{n=0}^\infty t^n \sum_{\lambda\vdash n} \frac{p_{\lambda}(x)}{z_\lambda}\,,
\end{equation}
where 
$p_\lambda(x)=\prod_{i=1}^k \left(\sum_j x_j^{l_i} \right)^{m_i}$ is the power sum symmetric polynomial corresponding to the partition $\lambda=(l_1^{m_i}\cdots l_k^{m_k})$. 
By setting $x_1=\cdots=x_M=1$ and $x_{M+1}=x_{M+2}=\cdots = 0$, we obtain  \eqref{eq:idpower}.

The terms of the order of ${\cal O}(1/N_c)$ in \eqref{eq:tmp2} merit scrutiny. 
This is because this expression contains power series of $N_f$ and $f_C(q,u)$, and if ${N_ff_C(q,u)}\sim {\cal O}(N_c)$, the contribution of these terms can be finite.
Hence, to dismiss the ${\mathcal{O}}(1/N_c)$ terms in equation (\ref{eq:tmp2}), it is necessary to consider the limit of $N_c \rightarrow \infty$ while maintaining $N_f f_C(q,u) \ll N_c$, equivalently $qN_f \ll N_c$, yielding the asymptotic form of the partition function:
\begin{equation}
  Z_G \to
  \left(\frac{2\pi}{\beta}\right)^{{N_f}{N_c}n_V}
  \left(1-(1-u)^2q^2\right)^{{N_f}{N_c}(n_E-n_V)}
  {\cal V}_G(q,u)^{N_fN_c} 
  \prod_{C\in [\Pi_+]}
  \frac{1}{\left(1-f_C(q,u)^2\right)^{N_f^2}}\,.
  \label{eq:Z1}
\end{equation}

\subsection{Evaluation by the saddle point approximation}

Next, we evaluate the unitary matrix integral of the partition function \eqref{eq:Zpre},
\begin{equation*}
  \int\prod_{e\in E}dU_e\,
  \zeta_G(q,u;U)^{N_f}\,,
\end{equation*}
in the limit of large $N_f$. 
Firstly, the gauge is fixed appropriately by utilizing the gauge invariance of $\zeta_G(q,u;U)$ and the left-right invariance of the Haar measure. 
This is achieved by setting $U_e=\bsone_{N_c}$ on the edges within a spanning tree of the connected graph $G$, resulting in $r = n_E - n_V + 1$ unitary matrices, which are the Wilson loops corresponding to 
the fundamental cycles of the graph $G$, 
remaining.
The edges out of the spanning tree are labelled as $\{e_a\mid a=1,\cdots,r\}$ and the remaining matrices are renamed as $U_a\equiv U_{e_a}$. Subsequently, the integral becomes
\begin{equation}
  \int\prod_{a=1}^r dU_a\,
  \zeta_G(q,u;U)^{N_f}
   \equiv 
  \int\prod_{a=1}^r dU_a\,
   e^{-N_f S[U]}\,.
   \label{eq:int2}
\end{equation}

At large $N_f$,  the integral \eqref{eq:int2} can be precisely evaluated through the application of the saddle-point approximation.
From the expression \eqref{eq:edge U-Bartholdi}, the ``action'' $S[U]$ can be written as
\begin{align}
  S[U] &= -\log \zeta_G(q,u;U) \nn \\
  &= \Tr\log\left(1-q (W_U + uJ_U)\right)\,.
  \label{eq:SU}
\end{align}
By defining 
\begin{equation}
  \delta A_a \equiv -i \delta U_a U_a^{-1}\,,
\end{equation}
the variation of the action can be represented as
\begin{equation}
  \delta S[U] = -i\sum_{a=1}^r \sum_{C\in R_a} f_C(q,u)\Tr\left( \delta A_a (U_C-U_C^\dagger) \right)\,,
\end{equation}
where 
$R_a$ constitutes the set of reduced but not necessarily primitive cycles that commence with $e_a$, and $f_C(q,u)$ designates a function of $q$ and $u$ specific to $C$, defined similarly as in \eqref{eq:VG and FC}.
To satisfy $\delta S[U]=0$ for any value of $q$ and $u$, it is necessary that
\begin{equation}
  U_C=U_C^\dagger
  \label{eq:cond_gen}
\end{equation}
be satisfied by all Wilson loops.
In particular, the same condition must be satisfied by 
the Wilson loops along the fundamental cycles:
\begin{equation}
  U_a^\dagger = U_a\,.
  \label{eq:cond_a}
\end{equation}
Since any Wilson loop is generated by $U_a$ 
($a=1,\cdots,r$), the conditions \eqref{eq:cond_gen} and \eqref{eq:cond_a} entail
\begin{equation}
  U_a U_b = (U_a U_b)^\dagger = U_b U_a\, 
\end{equation}
for ${}^\forall a, b\in\{1\cdots,r\}$. 
As a result, $U_a$ ($a=1,\cdots,r$) can be diagonalized simultaneously and the saddle points are given by
\begin{equation}
  U_a = {\rm diag}(\pm1,\cdots,\pm1) \,,
  \label{eq:saddles}
\end{equation}
modulo (global) unitary transformation. 

The value of the action at one of the saddle points is evaluated as 
\begin{align}
  S[U]\bigr|_{\rm FP} = -\sum_{C\in[{\cal P}_R]} \sum_{m=1}^\infty \frac{1}{m}f_C(q,u)^m \left(N_C^+  + N_C^- (-1)^m \right) + {\rm const.}\,,
\end{align}
where $N_C^\pm$ signifies the number of $\pm 1$ in the eigenvalues of $U_C$ on the saddle point and the constant term corresponds to the contribution from ${\cal V}_G(q,u)$.
The saddle point yielding the minimal action is
\begin{equation}
  U_a \equiv U_a^{(0)}= \bsone_{N_c}\,,
  \label{eq:vac}
\end{equation}
for all $a\in\{1,\cdots,r\}$ and the values of $e^{-N_f S[U]}$ at this point is expressed by the Bartholdi zeta function \eqref{eq:Bartholdi zeta} as
\begin{equation}
  e^{-N_f S[U^{(0)}]} = \zeta_G(q,u)^{N_f N_c}\,.
\end{equation}
Compared to it, the values of $e^{-N_f S[U]}$ at other saddle points are negligible in the order of $e^{-N_f}$ in the limit of large $N_f$.

However, it is important to note that the value of $N_c$ determines the validity of disregarding saddle points other than \eqref{eq:vac}. 
Since the number of the saddle points is of the order of $e^{N_c}$, if $N_c \gtrsim {\cal O}(N_f)$,
the total contribution from all saddle points cannot be dismissed even if each individual contribution is proportional to $e^{-N_f}$.
Therefore, in contrast to the previous subsection, 
we evaluate \eqref{eq:int2} in the region of $N_c \ll N_f$ and take the limit of $N_c\to\infty$ in the rest of this subsection.

In this region, it is only necessary to calculate the second derivative of the action \eqref{eq:SU} about the vacuum \eqref{eq:vac} to evaluate \eqref{eq:int2}: 
\begin{align}
  S[U] &= -N_c \log\zeta_G(q,u) -\sum_{a,b=1}^r \Tr_{N_c}\left(\delta A_a \delta A_b\right) ({\cal M}_G)_{ab} + {\cal O}(\delta A^3)\,,
\end{align}
where ${\cal M}_G$ is a square matrix of size $r$, which is comprised of the elements of the matrix appearing in the edge expression of the Bartholdi zeta function \eqref{eq:Bartholdi zeta}, 
\begin{equation}
  Y \equiv (\bsone_{2n_E} -q(W+uJ))^{-1}\,,
\end{equation}
as 
\begin{align}
  ({\cal M}_G)_{ab} = 2\left(-Y_{e_a e_a} \delta_{ab} + Y_{e_ae_b} Y_{e_be_a} - Y_{e_a e_b^{-1}}Y_{e_b^{-1} e_a}\right)\,.
  \label{eq:Mab}
\end{align}
Therefore, in the limit of $N_c\to\infty$ with $N_c\ll N_f$, the integral \eqref{eq:int2} can be exactly evaluated as 
\begin{align}
  \int\prod_{a=1}^r dU_a\,
  \zeta_G(q,u;U)^{N_f}
   &=
  \int\prod_{a=1}^r dU_a\,
   e^{-N_f S[U]} 
   = {\cal N} \zeta_G(q,u)^{N_fN_c} \left(\det{\cal M}_G\right)^{-\frac{N_c^2}{2}}\,,
\end{align}
with an irrelevant constant ${\cal N}$, 
and we see that the partition function \eqref{eq:Zpre} behaves asymptotically as
\begin{align}
  Z_G \to {\cal N} \left(\frac{2\pi}{\beta}\right)^{{N_f}{N_c}n_V}
  &\left(1-(1-u)^2q^2\right)^{{N_f}{N_c}(n_E-n_V)}
  {\cal V}_G(q,u)^{N_fN_c} \nn \\
  &\times \left(
  \prod_{C\in [\Pi_+]} \frac{1}{(1-f_C(q,u))^{2N_fN_c}} 
  \right)
  \left(\det{\cal M}_G\right)^{-\frac{N_c^2}{2}}\,.
  \label{eq:Z2}
\end{align}

\subsection{Phase structure in general graphs}

From \eqref{eq:Z1} and \eqref{eq:Z2}, we see that the free energy $F_G = -\frac{1}{N_c^2}\log Z_G$ of the model in large $N_c$ behaves as 
\begin{equation}
   F_G \to F_G^- \equiv  \gamma^2 \sum_{C\in[\Pi_+]} \log(1-f_C(q,u)^2) 
   \label{eq:FG-}
\end{equation}
in the region $q \ll \frac{1}{\gamma}$ and 
\begin{equation}
   F_G \to F_G^+ \equiv 2\gamma \sum_{C\in[\Pi_+]} \log(1-f_C(q,u)) + \frac{1}{2}\Tr\log{\cal M}_G+{\rm const.} 
   \label{eq:FG+}
\end{equation}
in the region $\gamma \gg 1$ 
up to common irrelevant terms. 
As already discussed, the FKM model on a graph with 
$r$ fundamental cycles 
is characterized by $r$ mutually independent unitary matrices. 
The qualitative discussion in the previous section remains applicable in this context: The distribution of eigenvalues of the unitary matrices is the result of the interplay between the measure-induced repulsive forces and the attractive forces derived from the potential term. Hence, when $N_f$ is small, the eigenvalues of all unitary matrices are dispersed along the circles due to the insufficient attractive force, conversely, when $N_f$ is substantial, the eigenvalues tend to cluster near the origins. This indicates that $F_G^-$ and $F_G^+$ correspond to the free energy in the phases where the eigenvalue densities of all unitary matrices are dispersed along the circles or concentrated near the origins, respectively. 
  
This can be understood more clearly by applying \eqref{eq:FG-} and \eqref{eq:FG+} to cycle graphs.
In fact, the cycle graph $G=C_n$ has only one 
fundamental cycle 
and the ``matrix'' ${\cal M}_{C_n}$ is  
\begin{equation}
  {\cal M}_{C_n} = \frac{2\alpha}{(1-\alpha)^2}\,,
\end{equation}
with $\alpha=f_C(q,u)$ for the unique 
fundamental cycle 
$C$. 
Therefore, the free energies \eqref{eq:FG-} and \eqref{eq:FG+} can be written as 
\begin{equation}
   F_{C_n}^- = \gamma^2 \log(1-\alpha^2) \,,
\end{equation}
and
\begin{equation}
   F_{C_n}^+ = (2\gamma-1) \log(1-\alpha) + \frac{1}{2}\log\alpha +{\rm const.}\,, 
   \label{eq:CnF+}
\end{equation}
respectively,  which reproduce the exact result \eqref{eq:FCn}.

In general, however, there are $2^r$ potentially possible phases that can be realized, dependent upon whether the eigenvalue distributions of the $r$ unitary matrices are dispersed or concentrated on the circles. 
The expressions \eqref{eq:FG-} and \eqref{eq:FG+} are believed to correspond to the free energies of the two most disparate phases, respectively.
As the parameter of the theory shifts from the phase with the free energy \eqref{eq:FG-}, where all eigenvalue distributions are dispersed, it is anticipated that the eigenvalue distributions of the $r$ unitary matrices will sequentially be separated, until the phase with the free energy \eqref{eq:FG+} is eventually realized.
In essence, in a general graph with $r>1$, we expect that the FKM model will have $r+1$ different phases,
with the GWW phase transitions occurring at their boundaries. 
Note that the order of these phase transitions is not necessarily third-order because, unlike the cycle graph discussed in the previous section, there is nontrivial interaction between the unitary matrices through the infinitely many primitive cycles in general.
Therefore, the two expressions of the free energy \eqref{eq:FG-} and \eqref{eq:FG+} will not be directly contiguous, 
and an attempt to connect them would appear as if a first-order phase transition were occurring.

For a graph with larger symmetry, it is natural to assume that multiple unitary matrices should behave in a similar manner. 
In this case, the intermediate phases should collapse and the number of transitions should be smaller than $r$ 
unless the symmetry of the graph is spontaneously broken. 
In particular, if the graph consists of repetitions of the same plaquette, such as a symmetric square lattice, 
all the intermediate phases will collapse and there should remain only two phases where the eigenvalues of all the unitary matrices are dispersed seamlessly around the circles as in (a) of Fig.~\ref{Distribution of Eigenvalue} or concentrated near the origins as in (b) of Fig.~\ref{Distribution of Eigenvalue}. 
This is the special case where it will make sense to discuss the phase transition between \eqref{eq:FG-} and \eqref{eq:FG+}.


In order to see it explicitly, let us consider the case of $u=0$ where the Bartholdi zeta function 
becomes
the Ihara zeta function. 
In this case, ${\cal V}_G(q,u)$ and $f_C(q,u)$ reduce to 
\begin{equation}
  {\cal V}_G(q,u=0) = 1, \quad 
  f_C(q,u=0) = q^{|C|}\,,
\end{equation}
and therefore \eqref{eq:FG-} and \eqref{eq:FG+} can be written through the Ihara zeta function $\zeta_G(q)$ as 
\begin{align}
  \begin{split}
  F_G^-\bigr|_{u=0} &= -\frac{\gamma^2}{2} \log \zeta_G(q^2)\,, \\
   F_G^+\bigr|_{u=0} &= -\gamma \log \zeta_G(q) + \frac{1}{2}\Tr\log{\cal M}_G + {\rm const.}\,,
  \end{split}
  \label{eq:FG+-}
\end{align}
which can be explicitly evaluated by using the determinant expression of the Ihara zeta function in \eqref{eq:Ihara zeta}. 

\begin{figure}[H]
  \begin{center}
  \includegraphics[scale=0.6]{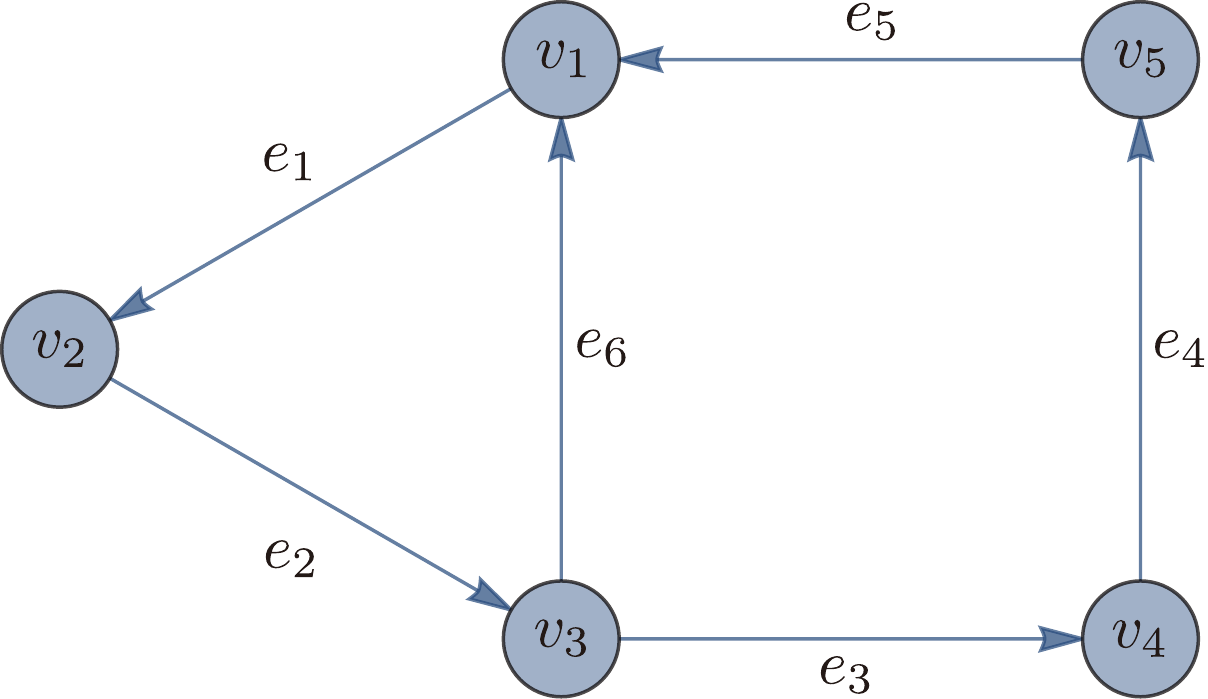}
  \end{center}
  \caption{A triangle-square (TS) graph with five vertices and six edges.}
  \label{fig:TS}
\end{figure}

\begin{figure}[H]
  \begin{center}
  \includegraphics[scale=0.45]{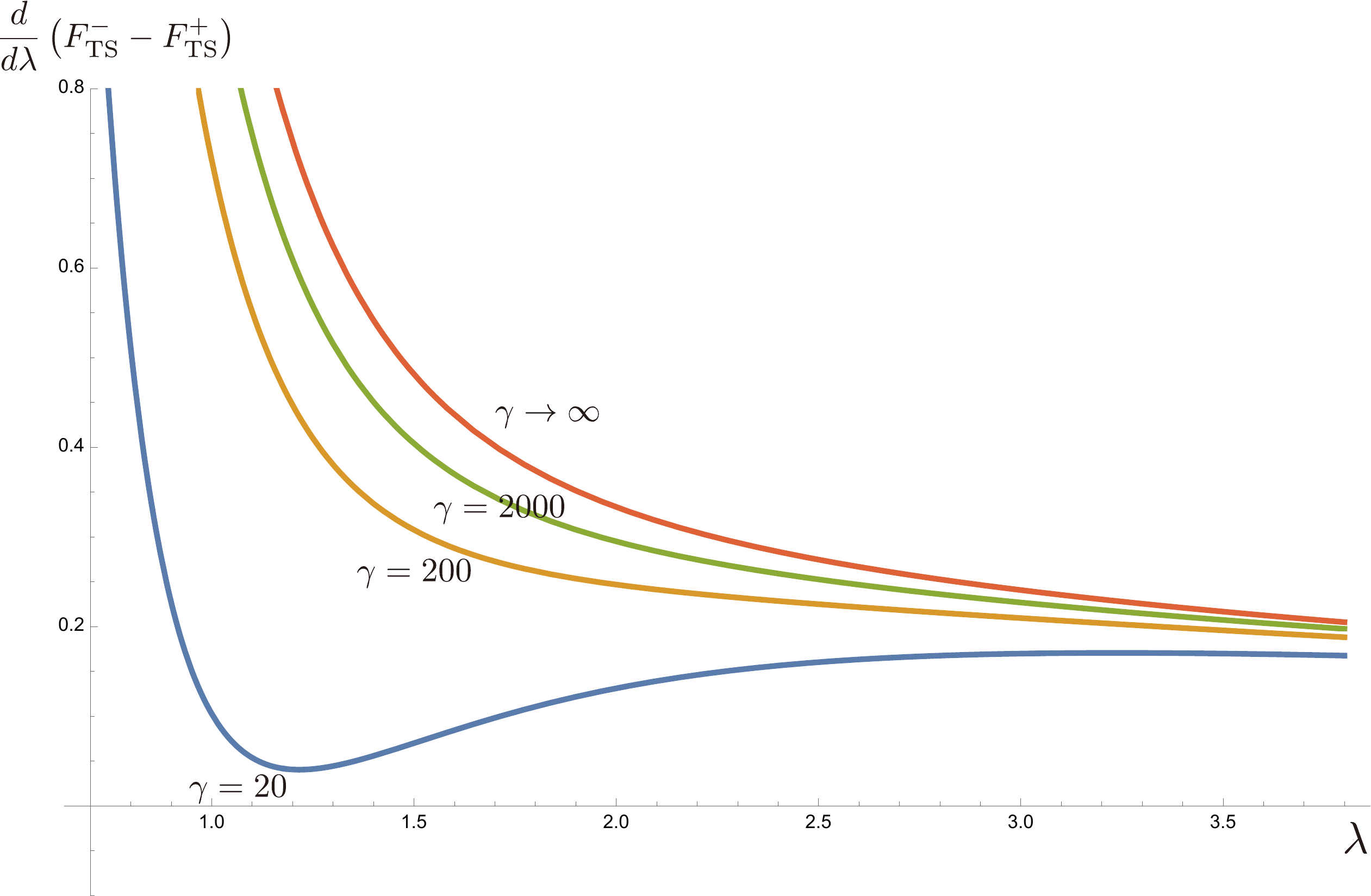}
  \end{center}
  \caption{For the triangle-square asymmetric graph, this plot shows the behavior of $\frac{d}{d\lambda}(F_{\rm TS}^- -F_{\rm TS}^+)$ for $\gamma=20$, $200$, $2000$
  and the $\gamma \to \infty$ limit in the region of the coupling $\lambda\sim {\cal O}(1)$, with setting $q=(\gamma\lambda)^{-1/3}$. 
  Since these curves do not cross nor touch the $\lambda$-axis,
$F_{\rm TS}^-$ and $F_{\rm TS}^+$ cannot be connected smoothly
in this region.
}
  \label{fig:F1-F2_TS}
\end{figure}

As an example of an asymmetric graph, we consider such a graph that a triangle and a square are joined by one edge, called TS.  (See Fig.~\ref{fig:TS}.)
The Ihara zeta function and $\det {\cal M}_G$ for $G={\rm TS}$ is evaluated as
\begin{equation}
    \begin{split}
        \zeta_{\rm TS}
        &=\frac{1}{(1-q^2)\left(1+q+q^2-q^4-2q^5\right) \left(1-q+q^2-2q^3+q^4-2q^5\right)}\,, \\
        \det {\cal M}_{\rm TS} 
        &=\frac{4 q^{7} \left(1+q+3q^2+q^3+3q^4+q^5+q^6\right)}{\left(1+q+q^2-q^4-2q^5\right)^{2} \left(1-q+q^2-2q^3+q^4-2q^5\right)^{2} }\,.
    \end{split}
\end{equation}
and the free energies $F_{\rm TS}^-$ and $F_{\rm TS}^+$ are obtained by substituting them into \eqref{eq:FG+-}. 
In order to compare the behavior of the free energies in the limit of \eqref{eq:limit}, we set $q=(\gamma \lambda)^{-1/3}$ and regard $\lambda$ as the coupling constant of the order of ${\cal O}(1)$%
\footnote{
Although we see a fluctuating behavior in the derivative of the free energy in the region of $|\lambda|\ll 1$ with finite $\gamma$, it is thought to be non-universal since this behavior vanishes in the limit of $\gamma\to\infty$. 
}.
Since it is difficult to compare the free energy directly due to the presence of the constant term in \eqref{eq:CnF+}, the first-order derivatives are compared instead.
We expect that they represent the free energies of two non-adjacent phases and thus they cannot be smoothly connected. 
Fig.~\ref{fig:F1-F2_TS} is the typical behavior of $\frac{d}{d\lambda}(F_{\rm TS}^- -F_{\rm TS}^+)$ for large $\gamma$. 
The curves never cross nor touch the $\lambda$-axis, which means that $F_{\rm TS}^-$ and $F_{\rm TS}^+$ cannot be connected smoothly as expected. 
It is natural to assume that there is another phase in the middle of these phases. 

\begin{figure}[H]
  \begin{center}
  \includegraphics[scale=0.6]{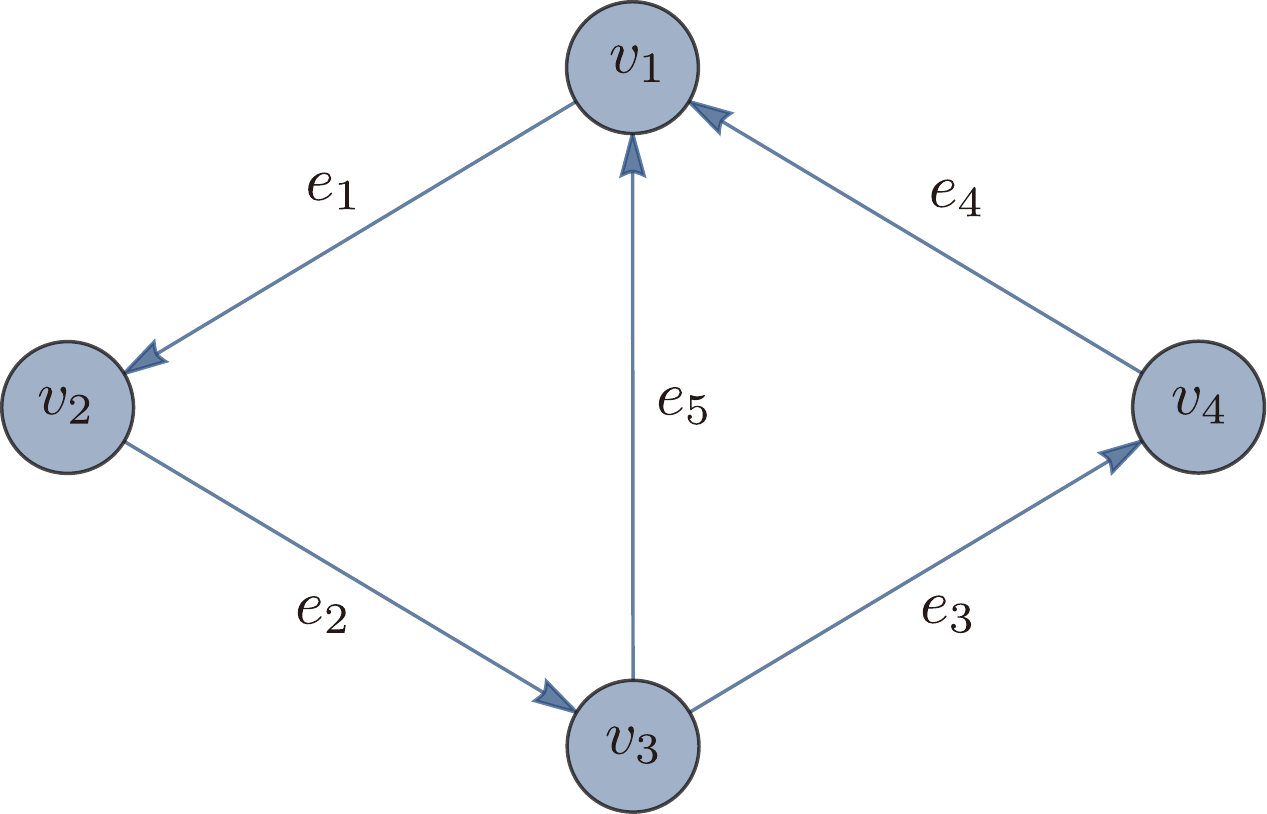}
  \end{center}
  \caption{A double triangle (DT) graph with four vertices and five edges.}
  \label{fig:DT}
\end{figure}

\begin{figure}[H]
  \begin{center}
  \includegraphics[scale=0.45]{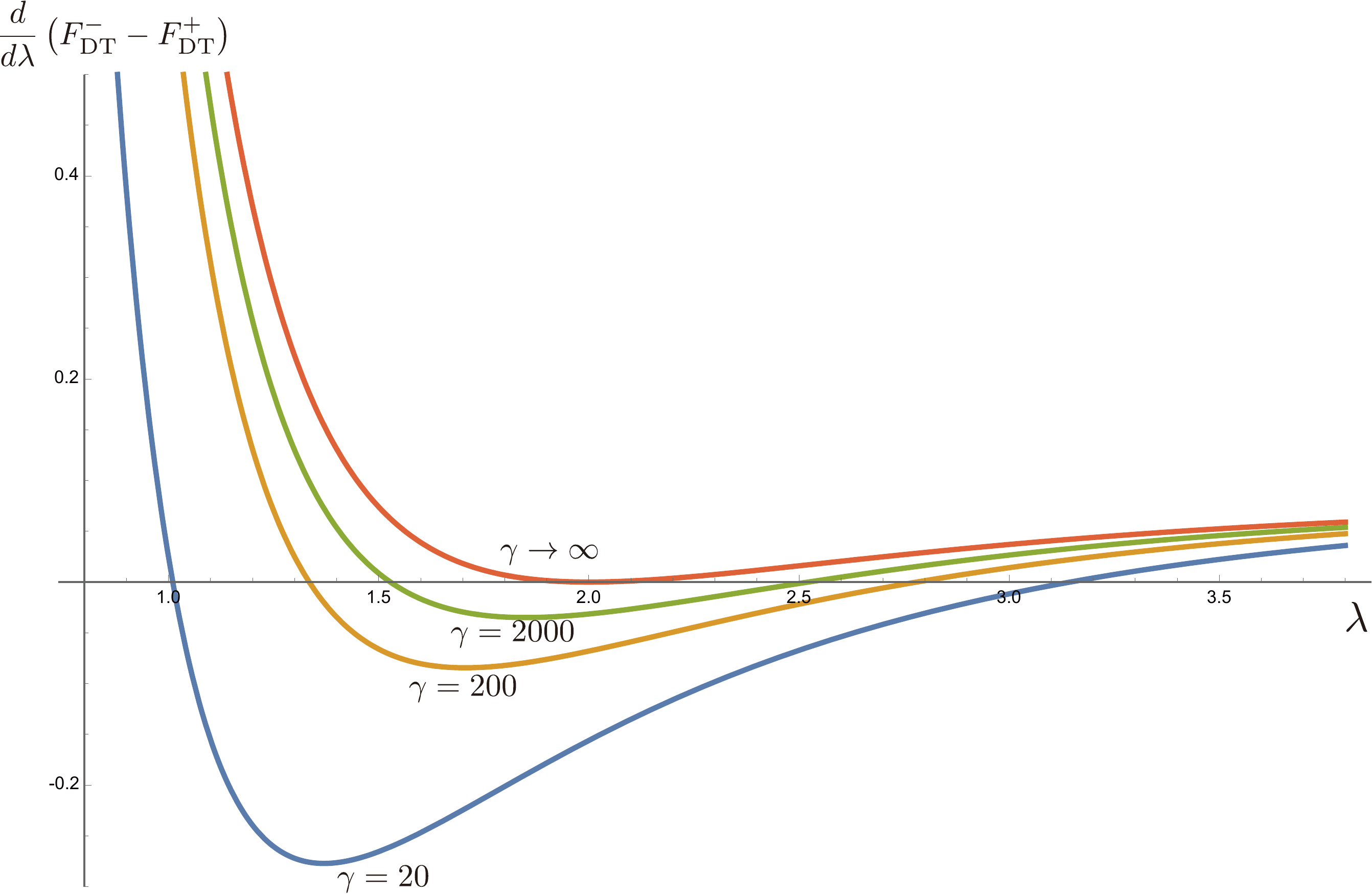}
  \end{center}
  \caption{For the double triangle (DT) graph, this plot shows that the behavior of $\frac{d}{d\lambda}(F_{\rm DT}^- -F_{\rm DT}^+)$ for $\gamma=20$, $200$, $2000$
  and the $\gamma \to \infty$ limit with setting $q=(\gamma\lambda)^{-1/3}$. We see that the third-order phase transition never occurs for the finite $\gamma$,
  since the corresponding curves cross the $\lambda$-axis.
  The curve for $\gamma\to\infty$ contacts with the $\lambda$-axis at $\lambda=2$,
  which is the third-order phase transition point of the GWW model.}
  \label{fig:F1-F2_DT}
\end{figure}

Next, as an example of a symmetric graph, let us evaluate \eqref{eq:FG-} and \eqref{eq:FG+} for a graph of two triangles pasted together by one edge, which we call the double triangle (DT).
(See Fig.~\ref{fig:DT}.)
The Ihara zeta function and $\det {\cal M}_G$ for $G={\rm DT}$ are given by 
\begin{align}
\begin{split}
\zeta_{\rm DT}(q) &= 
\frac{1}{ \left(1-q^4\right) \left(1+q^2-2q^3\right)\left(1-q^2-2q^3\right)}\,,
\\
  \det{\cal M}_{\rm DT} &=
  \frac{4 q^{6} \left(1+q\right)^{2}}{\left(1+q^{2}\right)\left(1+q^2-2q^3\right)^{2} \left(1-q^2-2 q^{3}\right)^{2} }\,,  
\end{split}
\label{eq:DTzeta_M}
\end{align}
and the free energies $F_{\rm DT}^-$ and $F_{\rm DT}^+$ are again obtained by substituting them into \eqref{eq:FG+-}. 
As in the TS case, we  set $q=(\gamma \lambda)^{-1/3}$ and regard $\lambda$ as the coupling constant. 
Fig.~\ref{fig:F1-F2_DT} is the behavior of $\frac{d}{d\lambda}(F_{\rm DT}^- -F_{\rm DT}^+)$ for large $\gamma$. 
In this case, we expect that the phases expressed by $F_{\rm DT}^-$ and $F_{\rm DT}^+$ are adjacent to each other. 
We see that the phase transition is at most second-order for $\gamma < \infty$
since the derivative of the free energy crosses the $\lambda$-axis, 
even though we cannot estimate the point of the phase transition exactly due to the existence of the constant term in the free energy%
\footnote{
Another possibility is that there is an intermediate phase and we cannot connect $F_{\rm DT}^\pm$ directly. In that case, it means that the symmetry of the graph should be spontaneously broken. }.
Remarkably, $\frac{d}{d\lambda}(F_{\rm DT}^- -F_{\rm DT}^+)$ becomes tangent to the $\lambda$-axis at $\lambda=2$ in the limit of $\gamma\to\infty$, namely, in the limit of \eqref{eq:limit}. 
In the next subsection, 
we see that this indicates that the transition indeed reduces to a third-order GWW phase transition in the limit of \eqref{eq:limit}
and the same phenomenon occurs in a wide class of graphs. 

\subsection{Free energies in the Wilson action limit}
\label{sec:GWW in limit}

Let us again consider the limit \eqref{eq:limit} 
where the effective action of the model becomes the standard Wilson action \eqref{eq:SWilson}. 
It is easy to see the first terms of  the free energies \eqref{eq:FG-} and \eqref{eq:FG+} become
\begin{align}
\begin{split}
    \gamma^2 \sum_{C\in[\Pi_+]}
    \log\left(1-f_C(q,u)^2\right) &\to - \frac{r'}{\lambda^2}, \\
    2\gamma \sum_{C\in[\Pi_+]} \log\left(1-f_C(q,u)\right) &\to -\frac{2r'}{\lambda} \,, 
\end{split}
\end{align}
respectively, 
where $r'$ is the number of the representative of cycles with the minimal length $l$. 
The term with $\log\det{\cal M}_G$ can be evaluated by considering the meaning of the matrix element \eqref{eq:Mab}:
The diagonal element $({\cal M}_G)_{aa}$ is (twice of) the summation of $f_C(q,u)$ for such reduced cycles $C$ that start from $e_a$. 
Therefore, in the limit of $q\to 0$, only the leading contribution $2q^{|C_a|}$ survives. 
On the other hand, the off-diagonal element $({\cal M}_G)_{ab}$ is the summation of $f_C(q,u)$ for such reduced cycles $C$ that start from $e_a$ and include $e_b$. 
Then the off-diagonal elements of ${\cal M}_G$ behave in higher order than ${\cal O}(q^{|C_a|})$ for any $a$ in the limit of $q\to 0$.
Therefore, the leading behavior of $\det{\cal M}_G$ in small $q$ is $2^r q^{\sum_{a=1}^r |C_a|}$. 
As a result, in the limit \eqref{eq:limit}, the free energies \eqref{eq:FG-} and \eqref{eq:FG+} converge to
\begin{align}
\begin{split}
    F_G^- &\to -\frac{r'}{\lambda^2}\,,  \\
    F_G^+ &\to -\frac{2r'}{\lambda} - \frac{1}{2}\left( \frac{1}{l} \sum_{a=1}^r |C_a| \right) \log \lambda + {\rm const.} \,,
\end{split}
\label{eq:limitF}
\end{align}
respectively. 
For the special case where 
all fundamental cycles $C_a$ are of the same length $l$
such as the square lattice, \eqref{eq:limitF} reduces to
\begin{align}
\begin{split}
    F_G^- &\to -\frac{r}{\lambda^2}\,,  \\
    F_G^+ &\to -\frac{2r}{\lambda} - \frac{r}{2}\log\lambda + {\rm const.}\,,
\end{split}
\end{align}
For example, when the graph is the double triangle, we can read $\det{\cal M}_{\rm DT} \sim 4q^6$ for small $q$ from \eqref{eq:DTzeta_M}, which is consistent with the result with $l=3$ and $r=2$.

It is no coincidence that this representation is equal to $r$ times the free energy of the cycle graph \eqref{eq:FCn-limit} in the same limit.
As seen in Sec.~\ref{sec:decomp}, we can decompose all the primitive reduced cycles in the evaluation of the unitary matrix integral at least in the phase with dispersed eigenvalue distribution. 
If we take the limit \eqref{eq:limit}, only the cycles of minimal length survive and it yields $r$ copies of cycle graphs effectively. 
We have seen that the FKM model on the cycle graph shows the third-order GWW phase transition and the transition point is $\lambda=2$ in the limit \eqref{eq:limit}. 
This strongly suggests that the phase transition point of the present graph is also at $\lambda=2$ and the third-order phase transition occurs universally in this class of graphs. 

In the case where the graph is a two-dimensional square lattice, this is consistent with the result that the effective action of the FKM model reduces to the Wilson action in the limit of \eqref{eq:limit} since the Wilson action on a two-dimensional square lattice is equivalent to the GWW model.
Notably, however, this result extends to arbitrary-dimensional lattice Yang-Mills theories represented by the Wilson action, owing to the sole assumption that all 
fundamental cycles 
are of the same length. 
This implies that the third-order GWW phase transition is present in large $N$ lattice gauge theories of arbitrary dimensions.

On the other hand,
if the 
fundamental cycles
$\{C_a\}$ contain cycles of length larger than $l$, no third-order phase transition occurs between $F_G^\pm$ because the coefficient of $\log\lambda$ changes.
This can be interpreted as the result that the multiple phases that arise due to the smaller symmetry of the graph survive in this limit as well. 
However, this result does not directly reflect the phase structure of the lattice gauge theory with the Wilson action on such a graph, since the contribution of cycles with large lengths disappears and the effective action \eqref{eq:Seff} does not coincide with the Wilson action on the graph under consideration in this limit. 

\section{Conclusion and Discussion}
\label{sec:Conclusion and Discussion}

In this paper, we have generalized the model proposed in \cite{Arefeva:1993ik} onto an arbitrary simple directed graph. 
This model, which 
is referred to as the FKM model, constitutes a variation of the KM model in which the adjoint scalar fields are replaced by $N_f$ fundamental scalar fields. It was demonstrated that the partition function of the FKM model can be expressed as a unitary matrix integral of a matrix-weighted Bartholdi zeta function.
The effective action of the FKM model is comprised of a sum of all Wilson loops, in a form analogous to that of the Wilson action in lattice gauge theory.
This is a consequence of the fact that this model, unlike the original KM model, has no extra local $U(1)$ symmetry.
We showed that, by taking a suitable scaling limit, the effective action converges to the standard Wilson action on the graph. 
We evaluated the free energy of the FKM model on a cycle graph (one plaquette) at large $N_c$ and showed that this model undergoes a third-order GWW phase transition exclusively when $N_f > N_c$.  
We also evaluated the free energy of the FKM model in the limiting regions of $q N_f\ll N_c$ and $N_f\gg N_c$ at large $N_c$ on a general graph.
The large $N$ decomposition of Wilson loops is applied for the former calculation, and the saddle point method for the latter. 
We argued that the FKM model on general graphs has multiple phases at large $N_c$ and that these two expressions are the free energy corresponding to the phases at both ends of the weakly and strongly coupled sides, respectively. 
When we consider a graph with larger symmetry like the square lattice, 
the intermediate phases collapse and a transition occurs between these two phases.
We evaluated a scaling limit of the free energies and showed that for graphs in which 
the fundamental cycles 
(all plaquettes in other words) consist of cycles of the same length, 
a third-order GWW phase transition occurs between them at that limit.
Since this condition holds for square lattices of arbitrary dimension, this implies that the third-order GWW phase transition occurs in large $N$ lattice gauge theories of arbitrary dimension.

The property that the Wilson action can be reproduced by taking the scaling limit of the FKM model makes this model intriguing 
because it would open up the possibility of investigating the Yang-Mills theory through the FKM model.
Although this result is obtained at the tree level, 
the non-perturbative analysis performed in the concrete graphs in the limit of large $N_c$ suggests that the same conclusion can be obtained in quantum theory. 
In addition, similar results are obtained in our ongoing preliminary numerical simulations (we will publish it soon).
From these observations, it seems certain that the FKM model does indeed contain a lattice gauge theory with the Wilson action.
The most important conclusion of this paper is that the FKM model undergoes the GWW phase transition at large $N_c$, regardless of the detail of the graph. 
As mentioned in Introduction, this transition is closely related to the confinement/deconfinement phase transition of QCD. 
This is because, 
in the process where the eigenvalues of the Polyakov loop transits from being localized in a part of the circumference to being uniformly distributed and the expectation value of the Polyakov loop vanishes, 
a phenomenon that the eigenvalue distributions join on the circumference always occurs
\cite{hanada2020partial,Hanada:2019czd,Watanabe:2020ufk,Hanada:2022wcq}.
Thus, considered in conjunction with our conclusion that the FKM model encompasses the Wilson action as a limit of the parameters, this result suggests that the Wilson lattice gauge theory shares the same properties.
Therefore, we can expect that the FKM model will be used to clarify the nature of the confinement/deconfinement transition in QCD.


Although the phase transition in this paper is considered at large $N_c$ with {\it fixing} a graph, 
it is natural to assume, as is the case with the original GWW phase transition, that these transitions are not mere lattice artifacts but are related to the large $N$ phase transitions observed in the continuum Yang-Mills theory in each dimension (See e.g. \cite{largeN_review} and references therein). 
The FKM model has the remarkable feature that its effective action is represented by the graph zeta function, 
which can be analytically connected to the complex plane at least in the region where $|q|$ and $|u|$ are sufficiently small and have properties similar to the famous Riemann zeta function. 
Although the properties of the matrix-weighted graph zeta function proposed in \cite{matsuura2022kazakov,matsuura2022graph} have not yet been fully investigated, 
its similarity to the Artin L-function and its connection to covering graphs have been suggested. If a more comprehensive analysis of properties can be conducted, it could lead to the creation of a novel analytical approach to probe the non-perturbative aspects of Yang-Mills theory.
A similar unitary matrix model to the FKM is also discussed in \cite{kimura2021universal,kimura2021unitary}
where the multicritical phase transition is suggested by using the Tracy-Widom distribution of the random partitions.
This implies that the Schur-Weyl duality and the random partitions can be an alternative way to analyze the FKM model. 
These present promising prospects for future research.

As stated in the primary text, we expect that the FKM model is anticipated to exhibit multiple phases at large $N_c$. 
In order to confirm it, a promising approach would be numerical simulations.
It is straightforward to carry out a Monte-Carlo analysis for  asymmetric graphs like Fig.~\ref{fig:TS}.
In addition to the large $N_c$ limit, it is also interesting to examine the continuous limit of the FKM model. 
If the FKM model indeed belongs to the same universality class as the Yang-Mills theory, the results of the Yang-Mills theory should be reflected in the continuous limit of the FKM model. Verifying this through numerical simulations constitutes a significant area of inquiry.

In \cite{matsuura2022kazakov,matsuura2022graph}, the partition function of the KM model on the cycle graph was computed at finite $N_c$, whereas the value at finite $N_c$ of the partition function of the FKM model on the cycle graph for arbitrary $N_f$ is not known. The only exception is when $N_f=1$ where we can carry out the unitary matrix integral exactly by using results in \cite{Puchala2017Symbolic} and the result is 
\begin{equation}
    Z_{C_1}^{N_f=1} = \left( \frac{2\pi}{\beta} \right)^{N_c} \frac{\xi_+(q,u)^{-N_c}}{1-q^2\xi_+(q,u)^{-2}}\,.
\end{equation}
Unfortunately, there is no known integral formula available when $N_f>1$, and current technology can only reach complicated forms involving the Weingarten functions. However, the evaluation of the unitary matrix integral undoubtedly changes at $N_f= N_c$  because the rank of the rectangular matrix $\Phi_v$ changes here.
As demonstrated, the phase transition occurs exclusively when $N_f > N_c$ for substantial values of $N_c$. It is anticipated that the background of this change in model properties at $N_f = N_c$ is related to its property at finite $N_c$. Clarification of this point would be of great interest.

The expressions of the free energy
\eqref{eq:FG-} and \eqref{eq:FG+}
with $u\ne 0$
are still formal unless we know how to compute $f_C(q,u)$ in a general graph. 
For \eqref{eq:FG+}, we can instead rewrite $F_G^+$ by using the Bartholdi zeta function as 
\begin{equation}
  F_{G}^+ = -\gamma \log \zeta_G(q,u) + \frac{1}{2}\Tr\log{\cal M}_G\,, 
\end{equation}
up to a constant term, which can be evaluated explicitly by using the determinant formulas to express the Bartholdi zeta function. 
However, it is still an open question if $F_G^-$ in \eqref{eq:FG-} can be expressed by appropriate graph zeta functions. 
It is the case for $u=0$ as shown in \eqref{eq:FG+-}, 
but it is still unclear for $u\ne 0$ because 
we cannot express the term $\prod_{C\in[\Pi_+]}\frac{1}{1-f_C(q,u)^2}$ by using known graph zeta functions. 
It would be useful to define a novel graph zeta function.
To this end, we recursively reduce all bumps contained in a primitive cycle $C$. The obtained cycle is a reduced cycle and is a positive power of a primitive reduced cycle $\hat{C}$. 
Calling this power $w(C)$, we can define a function by the Euler product with ``fermionic'' signature, 
\begin{equation}
  \tilde \zeta_G(q,u) \equiv \prod_{C\in[{\cal P}]} (1-(-1)^{w(C)}q^{|C|}u^{b(C)})^{-1}\,.  
  \label{eq:new zeta}
\end{equation}
If $(-1)^{w(C)}$ is omitted, it becomes the Bartholdi zeta function. 
Repeating the same computation to show $\zeta_G(q,u)={\cal V}(q,u)\prod_{C\in[{\cal P}_R]} (1-f_C(q,u))^{-1}$ (see \cite{matsuura2022graph}), 
we can show 
\begin{equation}
  \tilde \zeta_G(q,u) = {\cal V}_G(q,u)\prod_{C\in[{\cal P}_R]} \left(1+f_C(q,u)\right)^{-1} \,. 
\end{equation}
Therefore, we can write $F_G^-$ as 
\begin{equation}
  F_{G}^- = -\frac{\gamma^2}{4}\log \left(\zeta_G(q,u)\tilde\zeta_G(q,u) \right) \,,
\end{equation}
up to a constant term. 
If the function \eqref{eq:new zeta} has a determinant expression, not only we can express the free energy in both regions explicitly for $u\ne 0$, but also the function \eqref{eq:new zeta} would serve as an interesting new graph zeta function. 
A richer relationship may be found between lattice gauge theory and the graph zeta function by exploring this possibility.

\section*{Acknowledgments}
The authors would like to thank 
M.~Hanada, M.~Honda and H.~Watanabe 
for helpful comments on the phase transitions in gauge theory and for providing us with references.
They would also like to thank 
A.~Flachi, 
T.~Fujimori,
T.~Ishii,
Y.~Hatsuda,
S.~Kanno, 
Y.~Nakayama, and
Y.~Yoshida
for useful discussions.
This work is supported in part
by Grant-in-Aid for Scientific Research (KAKENHI) (C), 
Grant Number 20K03934 (S.~M.) and Grant Number 23K03423 (K.~O.).

\appendix
\section{Graph zeta functions}
\label{app:zeta}

In this appendix, we summarize the definition and some properties of the Ihara and Bartholdi zeta functions. For more detail, see \cite{matsuura2022graph,matsuura2022kazakov}.

\subsection{Ihara and Bartholdi zeta functions}
\label{app:IandB}

The graph zeta functions are defined by the Euler product associated with primitive cycles of a given graph. 
The simplest graph zeta function is called the Ihara zeta function \cite{Ihara:original,MR607504,sunada1986functions} which is defined as the Euler product over the equivalence classes of primitive reduced cycles of a given graph $G$ 
\begin{equation}
  \zeta_G(q) = \prod_{C\in [{\cal P}_R]} \frac{1}{1-q^{|C|}}\,. 
  \label{eq:Ihara}
\end{equation}
The Ihara zeta function counts only reduced cycles, 
whereas the Bartholdi zeta function counts all cycles including bumps
\cite{bartholdi2000counting}: %
\footnote{See also \cite{mizuno2003bartholdi} for a generalization.}.
\begin{equation}
  \zeta_G(q,u) = \prod_{C\in [{\cal P}]} \frac{1}{1-q^{|C|} u^{b(C)}}\,.
  \label{eq:Bartholdi}
\end{equation}
We concentrate on the Bartholdi zeta function in the following since the Bartholdi zeta function reduces to the Ihara zeta function by setting $u=0$: 
\[
 \zeta_G(q,u=0)=\zeta_G(q)\,.
\]

It is remarkable that the Bartholdi zeta function (and thus also the Ihara zeta function) is represented as the inverse of a polynomial, even though the graph generally has infinitely many equivalence classes of cycles.
Let us define $2n_E\times 2n_E$ matrices 
$W$ and $J$ whose elements are defined by 
\begin{align}
    W_{\bse\bse'} = \begin{cases}
      1 & {\rm if}\ t(\bse) = s(\bse')\ {\rm and}\ \bse'^{-1}\ne \bse \\
      0 & {\rm others}
    \end{cases}\,,\quad
    J_{\bse\bse'} = \begin{cases}
      1 & {\rm if}\ \bse'^{-1}= \bse \\
      0 & {\rm others}
    \end{cases}\,,\quad
    \label{eq:WJ}
\end{align}
where $\bse, \bse' \in E_D$,
which are called the edge adjacency matrix and the bump matrix, respectively.
Using these matrices, the Bartholdi zeta function can be expressed as 
\begin{equation}
\zeta_G(q,u) = \det \left(\bsone_{2n_E}- q(W+uJ)\right)^{-1}\,,
\label{eq:edge Bartholdi}
\end{equation}
which is the inverse of a polynomial of $q$ and $u$ as announced.

The expression \eqref{eq:edge Bartholdi} is sometimes called the edge Bartholdi zeta function because it is described through matrices $W$ and $J$ that characterize the relation among the edges of the graph.
Apart from this expression, there is another expression focusing on the relation among the vertexes; 
\begin{equation}
\zeta_G(q,u) = \bigl(\bsone_{n_V}-(1-u)^2q^2\bigr)^{-(n_E-n_V)}
\det\bigl(\bsone_{n_V}-q A + (1-u)q^2(D-(1-u) \bsone_{n_V})\bigr)^{-1}\,,
\label{eq:vertex Bartholdi}
\end{equation}
where $D$ is the diagonal matrix defined by
$D={\rm diag}_{v\in V}(\deg(v))$ and
the matrix $A$ is a square matrix of size $n_V$ called the vertex adjacency matrix
defined by
\begin{equation}
A_{vv'} = \sum_{\bse\in E_D}
\delta_{\langle{v},{v'}\rangle,\bse} \,.  \quad (v,v'\in V). 
\label{eq:vertex A}
\end{equation}

\subsection{Matrix-weighted Bartholdi zeta function}
\label{app:matrix zeta}

We here consider placing an invertible $K\times K$ matrix $X_e$ to each edge $e$ of the graph.
We assume that the matrix on the inverse edge $e^{-1}$ is the inverse of the matrix on the edge $e$;
\begin{equation}
  X_{e^{-1}}=X_e^{-1}\,.
  \label{eq:matrix assumption}
\end{equation}
For a cycle $C=(\bse_{i_1}\cdots\bse_{i_n})$, 
we assign a matrix,
\[
X_C \equiv X_{\bse_{i_1}}\cdots X_{\bse_{i_n}}.
\]

The matrix-weighted Bartholdi zeta function \cite{matsuura2022kazakov} is defined by the Euler product as
\begin{align}
  \zeta_G(q,u;X) &\equiv \prod_{C\in [{\cal P}]} \det\bigl( \bsone_{K} - q^{|C|} u^{b(C)} X_C \bigr)^{-1}\,.
  \label{eq:W-Bartholdi}
\end{align}
Like the original Bartholdi zeta function,
the matrix-weighted Bartholdi zeta function \eqref{eq:W-Bartholdi} can be expressed as the inverse of the determinant of a matrix. 
This is achieved by extending the edge adjacency matrix
and the bump matrix \eqref{eq:WJ} as
\begin{align}
    (W_X)_{\bse\bse'} = \begin{cases}
    X_{\bse} & {\rm if}\ t(\bse) = s(\bse')\ {\rm and}\ \bse'^{-1}\ne \bse \\
      0 & {\rm others}
    \end{cases}\,,\quad
    (J_X)_{\bse\bse'} = \begin{cases}
      X_{\bse} & {\rm if}\ \bse'^{-1}= \bse \\
      0 & {\rm others}
    \end{cases}\,.
    \label{eq:WJw}
\end{align}
We can show that the matrix-weighted Bartholdi zeta function can be written as
\begin{align}
  \zeta_G(q,u;X)
  &= \det\bigl( \bsone_{2Kn_E} - q(W_X + u J_X)\bigr)^{-1}\,. 
  \label{eq:edge W-Bartholdi}
\end{align}

We can further show that 
the matrix-weighted Bartholdi zeta function can be expressed through the matrix-weighted vertex adjacency matrix of the size $K n_V$,
\begin{equation}
(A_X)_{vv'} = \sum_{\bse\in E_D} X_{\bse}\,
 \delta_{\langle{v},{v'}\rangle,\bse} \,,
 \label{eq:vertex Amat}
\end{equation}
as
\begin{equation}
\zeta_G(q,u;X) = \bigl(1-(1-u)^2q^2\bigr)^{-K(n_E-n_V)}
\det\bigl(\bsone_{Kn_V}-q A_X + (1-u)q^2(D-(1-u) \bsone_{Kn_V})\bigr)^{-1}\,,
\label{eq:vertex W-Bartholdi}
\end{equation}
where $D$ has been redefined as
\begin{equation}
  D_{vi;v'j} = \deg(v) \delta_{vv'} \delta_{ij}\,.
  \quad (i,j=1,\cdots,K)
  \label{eq:matD}
\end{equation}
Note that the equivalence between \eqref{eq:edge W-Bartholdi} and \eqref{eq:vertex W-Bartholdi} is spoiled if the condition \eqref{eq:matrix assumption} is not satisfied.

\subsection{Reduction of the cycles}
\label{app:reduced expression}

When a cycle has backtracking as $\tC=P_1 \bse \bse^{-1} P_2$, the matrix $X_\tC$ can be rewritten as 
$X_\tC=X_{P_1} X_{\bse} X_{\bse^{-1}} X_{P_2} = X_{P_1}X_{P_2}$ because we have assumed $X_{\bse^{-1}}=X_\bse^{-1}$. 
Therefore, if a cycle $\tC$ becomes a reduced cycle $C$ by reducing all the bumps repeatedly, we see the matrices on the cycles $\tC$ and $C$ are identical; $X_\tC=X_C$. 
Noting that a reduced cycle is a positive power of a primitive reduced cycle $C$ by definition, we denote the set of representatives of primitive cycles that are equivalent to $C^k$ $(k\in\N)$ after eliminating the bumps by $[{\cal B}(C^k)] \subset [{\cal P}]$. 
We also denote the set of the representatives of primitive cycles that reduce to a point (vertex) by eliminating the bumps by $[{\cal B}_0] \subset [{\cal P}]$.
Using this notation, we can rewrite \eqref{eq:W-Bartholdi} as
\begin{align}
  \zeta_G(q,u;X) &= {\cal V}_G(q,u)^K
  \prod_{C\in [{\cal P}_R]}
  \prod_{k=1}^\infty
  \prod_{\tC\in [{\cal B}(C^k)]}
  \det\bigl(\bsone_K- q^{|\tC|}u^{b(\tC)}X_{C}^k \bigr)^{-1} \nn \\ 
  &= {\cal V}_G(q,u)^K
  \prod_{C\in [{\cal P}_R]}
  \exp\left(
    \sum_{n=1}^\infty \frac{1}{n}{f_{C}(q,u)^n}
     \Tr(X_C^{n}) \right)\,,
  \label{eq:path W-Bartholdi}
\end{align}
where 
\begin{equation}
  f_{C}(q,u) \equiv \sum_{\tC\in [\cB(C)]} q^{|\tC|}u^{b(\tC)}\,,
  \label{eq:FC}
\end{equation}
and
\begin{align}
{\cal V}_G(q,u) 
&\equiv \prod_{\tC\in [{\cal B}_0]} \frac{1}{1-q^{|\tC|}u^{b(\tC)}}\,.
\label{eq:VG}
\end{align}
We note that the matrix-weighted Bartholdi zeta function can be further rewritten as
\begin{equation}
  \zeta_G(q,u;X) = {\cal V}_G(q,u)^K
  \prod_{C\in [{\cal P}_R]}
  \det\left(\bsone_K-f_C(q,u)X_C\right)^{-1}\,,\nn
\end{equation}
which can be regarded as an extension of the Ihara zeta function in the sense that we count rather $f_C(q,u)$ not $q^{|C|}$ for a primitive reduced cycle $C$.

\bibliographystyle{unsrt}
\bibliography{refs}

\end{document}